\pdfoutput=1

\documentclass[%
pra,twocolumn,showpacs,superscriptaddress,floatfix
]{revtex4-1}

\usepackage[dvipdfmx,dvipdfmx]{graphicx}
\usepackage{graphicx}
\usepackage{color}
\usepackage{siunitx}

\usepackage{subfigure}

\usepackage{epsfig}
\usepackage{float}
\usepackage{epstopdf}
\usepackage{url}

\usepackage{amsmath}
\usepackage{amsfonts}
\newcommand{\abs}[1]{\left| #1 \right|}
\newcommand{\ket}[1]{\left | #1 \right \rangle}
\def\k(#1){|#1\rangle}
\newcommand{\bra}[1]{\left \langle #1 \right |}

\newcommand{\beq}{\begin{equation}}
\newcommand{\eeq}{\end{equation}}
\newcommand{\beqa}{\begin{eqnarray}}
\newcommand{\eeqa}{\end{eqnarray}}
\newcommand{\beqan}{\begin{eqnarray*}}
\newcommand{\eeqan}{\end{eqnarray*}}

\newcommand{\affA}{%
\affiliation{
 Center for Macroscopic Quantum States (bigQ), Department of Physics, Technical University of Denmark, Building 307, Fysikvej, 2800 Kgs.~Lyngby, Denmark}
     }
\newcommand{\affB}{%
\affiliation{
Advanced ICT Research Institute, National Institute of Information and Communications Technology, 588-2, Iwaoka, Nishi-ku, Kobe, Hyogo 651-2492, Japan }}

\newcommand{\affC}{%
\affiliation{
Graduate School of Engineering, Kobe University, 1-1 Rokko-dai cho, Nada-ku, Kobe 657-0013, Japan}}

\begin{document}

\title{
Experimental demonstration of a quantum receiver beating the standard quantum limit at telecom wavelength
}


\author{Shuro Izumi}
\affA
\author{Jonas S. Neergaard-Nielsen}
\affA
\author{Shigehito Miki }
\affB
\affC
\author{Hirotaka Terai}
\affB
\author{Ulrik L. Andersen}
\affA

\begin{abstract}
Discrimination of coherent states beyond the standard quantum limit (SQL) is an important task not only for quantum information processing but also for optical coherent communication.
In order to optimize long distance optical fiber networks,
it is of practical importance to develop a quantum receiver beating the SQL and approaching the quantum bound at telecom wavelength.
In this paper, we experimentally demonstrate a receiver beating the conventional SQL at telecom wavelength.
Our receiver is composed of a displacement operation, a single photon counter and a real time adaptive feedback operation.
By using a high performance single photon detector operating at the telecom wavelength, 
we achieve a discrimination error beyond the SQL. The demonstration in the telecom band is an enabler for quantum and classical communication beyond the SQL using a coherent state alphabet, and we envision that the technology can be used for long-distance quantum key distribution, effective quantum state preparation and quantum estimation.   
\end{abstract}

\maketitle

\section{Introduction}\label{Sect:1}
Coherent states are inherently non-orthogonal and can therefore not be perfectly discriminated \cite{Helstrom_book76_QDET}.
This non-orthogonality is an attractive feature of coherent states, as it for example ensures the security of quantum key distribution \cite{Bennett,RevModPhys.74.145}. On the other hand, for classical optical coherent communications with a lossy channel such as satellite-to-ground laser communication \cite{4063386,LiaoQKD} and long distance optical fiber communication \cite{7174950,TakesueQKD}, the non-orthogonality sets a fundamental limitation on the attainable communication rate and distance. 
In long distance coherent fiber communication, 
optical repeaters, e.g. erbium-doped fiber amplifiers, are commonly used to amplify weak, attenuated signals.
Although these techniques are well established, additional noise in the amplification process limits the communication distance if a cascade of amplifiers are required to transmit a signal \cite{PhysRevD.26.1817}.
To some extent, this may be remedied by noiseless amplification with phase sensitive amplifiers \cite{Tong2011, Asobe2018PhaseSA}.
Another promising and important direction for further improvement of optical coherent communication is designing unconventional receivers detecting the transmitted signal states below the shot noise limit \cite{Helstrom_book76_QDET}.
Unlike the strategy relying on amplifiers, that enhances the signal to noise ratio for noisy channel communication, the receivers beating the shot noise limit enables us to attain or closely approach the ultimate performance set by quantum mechanics.
Furthermore, the receivers that can directly project an input optical state onto a desired basis are important resources for optical quantum information processing \cite{PhysRevA.68.042319}.
The efficiency of reading out the information of an encoded signal state is 
often studied in the context of quantum state discrimination where the discrimination error of possible candidate signal states is one of the relevant figures of merit.

If one performs direct detection of physical variables that are encoded in coherent states, the obtainable discrimination error is limited by the shot noise.
This error probability given by conventional detection techniques is defined as the standard quantum limit (SQL).
In quantum mechanics, however, measurements can be mathematically represented as positive operator valued measures (POVM) and the ultimate bound for the discrimination error is obtained by optimizing the POVMs \cite{Helstrom_book76_QDET}.
The minimum discrimination error is called the Helstrom bound which can be derived analytically for particular types of states \cite{915636} but a physical implementation of the optimal measurements is often non-trivial.
Receivers beating the SQL, which we call quantum receivers hereinafter, have been extensively explored both theoretically \cite{Kennedy73,Dolinar73,PhysRevA.54.2728,PhysRevA.78.022320} and experimentally \cite{PhysRevLett.101.210501,Tsujino:10,PhysRevLett.106.250503,PhysRevLett.121.023603,CookMartinGeremia2007_Nature}.
For binary phase shift keying signals (BPSK),
performance beyond the SQL is achievable with a simple detection strategy consisting of a displacement operation followed by photon counting \cite{PhysRevLett.101.210501,Tsujino:10,PhysRevLett.106.250503,PhysRevLett.121.023603}.
Furthermore, the Helstrom bound can be attained by introducing an adaptive feedback operation that optimizes the displacement operation depending on the outcomes of the photon counter \cite{Dolinar73,CookMartinGeremia2007_Nature}.
While this feedback receiver was originally proposed for the BPSK coherent states discrimination,
it has been shown that the receiver can implement arbitrary two-dimensional projective measurements \cite{TakeokaSasakiLutkenhaus2005,TakeokaSasakiLutkenhaus2006_PRL_BinaryProjMmt}.
Recently, the displacement based photon detection receivers with or without feedback have been used to realize a single-rail qubit projector which was characterized by quantum detector tomography \cite{Kennedytomography,PhysRevLett.124.070502}.

In addition to binary state discrimination,
various types of quantum receivers have been developed for quaternary phase shift keying (QPSK) coherent states \cite{Bondurant93,M_ller_2012,izumi2012,izumi2013,Becerra13,Becerra15,Ferdinand}.
A significant improvement of the discrimination error beyond the  SQL has been experimentally observed by optimizing the feedback strategy according to the {\it a posteriori} probability for observed events \cite{Becerra13}.
Moreover, a practical receiver with photon number resolving detector, that is robust against mode matching imperfection of the displacement and dark count noise, has been analyzed \cite{izumi2013, 6600857} and demonstrated \cite{Becerra15}.
At the telecom wavelengths, however, implementation of quantum receivers that operate beyond the SQL has been particularly challenging mainly due to the low efficiency and high dark count rates of conventional telecom based photon counters \cite{PhysRevA.101.032306}. 

In this paper,
we experimentally demonstrate an all-fiber-based quantum receiver at telecom wavelength that beats the SQL for the discrimination of QPSK states.
Our receiver consists of a displacement operation, photon detection and feedback for updating the displacement.
For the photon detection, we employ a superconducting nanowire single photon detector which shows high performance at the telecom wavelength \cite{SSPD,SSPD2,Marsili}.
Our receiver achieves more than $65\%$ detection efficiency thereby allowing for operation beyond the SQL without compensating for any imperfections.
We further investigate the effect of feedback delay in the receiver by developing a mathematical model and we
find that the delay can significantly degrade the performance of the receiver for large signals.

Our paper is organized as follows.
We first introduce the quantum receiver and derive the theoretically achievable error probability (Sec.~\ref{Sect:2}).
In Sec.~\ref{Sect:3},
we discuss our experimental setup and experimental results while concluding in
Sec.~\ref{Sect:4}.
\section{Quantum receiver with feedback for QPSK signals}\label{Sect:2}
\begin{figure}[b]
\centering 
{
\includegraphics[width=1.0\linewidth]
{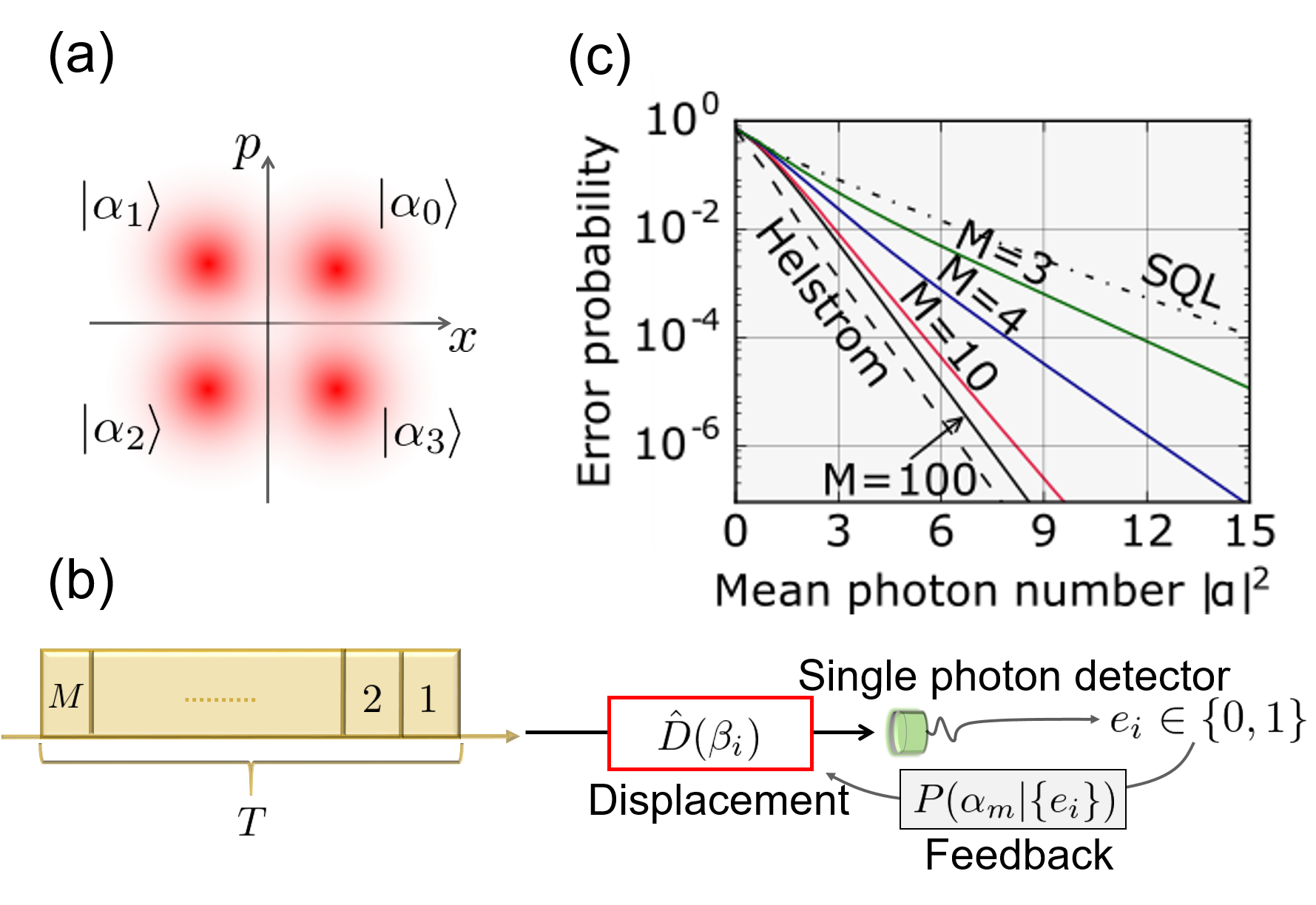}
}
\caption{
(a) QPSK coherent states in phase space.
(b) Schematic of a receiver with displacement operation, photon detection and feedback operations. 
(c) Theoretical performance of the receiver. Black, red, blue and green solid lines represent the receiver with $M=100$, $M=10$, $M=4$ and $M=3$.
Dashed-dot and dashed lines are the standard quantum limit and the Helstrom bound.
\label{Dolinar_config}
}
\end{figure}
In this section, we introduce a quantum receiver beating the SQL and approaching the Helstrom bound for the discrimination of the QPSK coherent states.

We define the QPSK coherent states as
$\ket{\alpha_m}=\ket{\abs{\alpha} e^{(2m+1) i\pi/4}}$ where $m=0,1,2,3$ and $\abs{\alpha}$ represents the magnitude of the signal state (Fig.~\ref{Dolinar_config}(a)).
Figure \ref{Dolinar_config}(b) depicts a schematic of the receiver.
The receiver consists of a displacement operation, a single photon detector (SPD) and real-time feedback control of the displacement phase dependent on the counting history of the SPD.
The displacement operation
can be physically implemented by combining the signal state with a strong reference beam at a beam splitter with nearly unit transmittance.
To illustrate the feedback control of the displacement, an incoming signal state with full time width $T$ is virtually divided into $M$ temporal mode bins. 
At each step, $i$, of the protocol, a displacement is applied to the $i$th temporal mode such that one of the candidate states is displaced to the vacuum state.
The SPD provides a binary outcome $e_i\in\{0( \mathrm{off}),1(\mathrm{on})\}$ whose POVM is respectively represented by $\hat{\Pi}_{0}=e^{-\nu}
\sum_{n=0}^{\infty} (1-\eta)^n\ket{n}\bra{n}$, $\hat{\Pi}_{1}=\hat{I}-\hat{\Pi}_{0}$, where $\nu$ is the dark count noise (counts/state) and $\eta$ the detection efficiency of the SPD \cite{BARNETT199845}.
The probability of getting an outcome $e_i$ for a coherent state $\ket{\gamma}$ with a displacement operation $\hat{D}(\beta)=\exp(\beta \hat{a}^{\dagger} - \beta^{\ast} \hat{a})$ is therefore given by
\begin{eqnarray}
p(e_i|\gamma;\beta)&=&
\bra{\gamma}\hat{D}^{\dagger}(-\beta)\hat{\Pi}_{e_i}\hat{D}(-\beta)\ket{\gamma} 
\nonumber
\\
&=&
(1-e_i)e^{-\nu-\eta\abs{\gamma-\beta}^2}
\nonumber
\\
&+&
e_i(1-e^{-\nu-\eta\abs{\gamma-\beta}^2}).
\end{eqnarray}

The {\it a priori} belief was that the incoming state was the one that was displaced to the vacuum.
If the detector outcome is ``off'', this belief is reinforced and the same displacement is maintained for the next step. If, on the other hand, the outcome is ``on'', this is a strong indication that the incoming state was {\it not} displaced to the vacuum. Hence, the displacement phase should (typically) be changed for the next step. Based on the full photon detection history $\{e_r\}$, the candidate state which has the highest {\it a posteriori} probability $P(\alpha_m|\{e_i\})$ should now be displaced towards the vacuum state \cite{Becerra13}.

The {\it a posteriori} probability after detecting the $j$'th bin is obtained from
\begin{eqnarray}
P(\alpha_m|\{e_i\})=\frac{\Pi_{k=1}^{j} p(e_k|\alpha_{m}/\sqrt{M};\beta_k)}{\sum_{l=0}^{3}\Pi_{k=1}^{j} p(e_k|\alpha_{l}/\sqrt{M};\beta_k)},
\label{eqposteriori}
\end{eqnarray}
where we assume that the {\it a priori} probabilities of the four signal states are equal and the signal state is equally divided into $M$ temporal modes.
The most likely signal state that maximizes the final {\it a posteriori} probability is concluded to be the received state.
Thus, the average error probability for the discrimination of the QPSK coherent states
using the feedback receiver is represented as
\begin{equation}
P_e=1-\frac{1}{4}\sum_{m=0}^3\sum_{\{e_i\}\in O_m}  P(\{e_i\}|\alpha_{m}),
\label{P_e}
\end{equation}
where the conditional probability $P(\{e_i\}|\alpha_m)$ represents the probability of having the outcomes $\{e_i\}$ if the actual incoming signal is $\ket{\alpha_m}$ and $O_m$ is the set of outcomes for which it was concluded that the state $\ket{\alpha_m}$ was received.
An analytical expression for Eq.~(\ref{P_e}) was obtained for the ideal case \cite{izumi2012}.
On the other hand, finding a solution for the general case with experimental imperfections is not straightforward and we thus evaluate the error probability with imperfections by simulating the model.

The achievable error probability with the receiver in the ideal condition is shown in Fig.~\ref{Dolinar_config}(c). 
The ultimate bound for the discrimination of the QPSK states is given by the Helstrom bound \cite{Helstrom_book76_QDET, izumi2012}.
The SQL is the error probability that can be attained by the conventional heterodyne receiver measuring both quadratures, whose error probability is calculated to be,
\begin{eqnarray}
P_e^{\mathrm{SQL}}=1-\frac{1}{4}\bigl(1+\mathrm{erf}(\abs{\alpha}/\sqrt{2})\bigr)^2,
\end{eqnarray}
where $\mathrm{erf}(x)$ is the error function defined as
$\mathrm{erf}(x)=\frac{2}{\sqrt{\pi}}\int_0^x \mathrm{e}^{-t^2} dt$.
Under ideal conditions, already the minimum number of feedback steps $M=3$ beats the SQL.
Moreover, by increasing the number of feedback steps up to e.g. $M=10$, the feedback measurement closely approaches the Helstrom bound. 
Though the analytical form for the asymptotic limit of the number of feedback steps $M\rightarrow\infty$ is not straightforward,
the improvement of the error probability saturates for large numbers of steps.
In contrast to the BPSK case, there is a small gap between the Helstrom bound and the feedback measurement based on the photon detection \cite{izumi2012}.

\section{Experiment}\label{Sect:3}

\begin{figure}[t]
\centering
{
\includegraphics[width=1.0 \linewidth]
{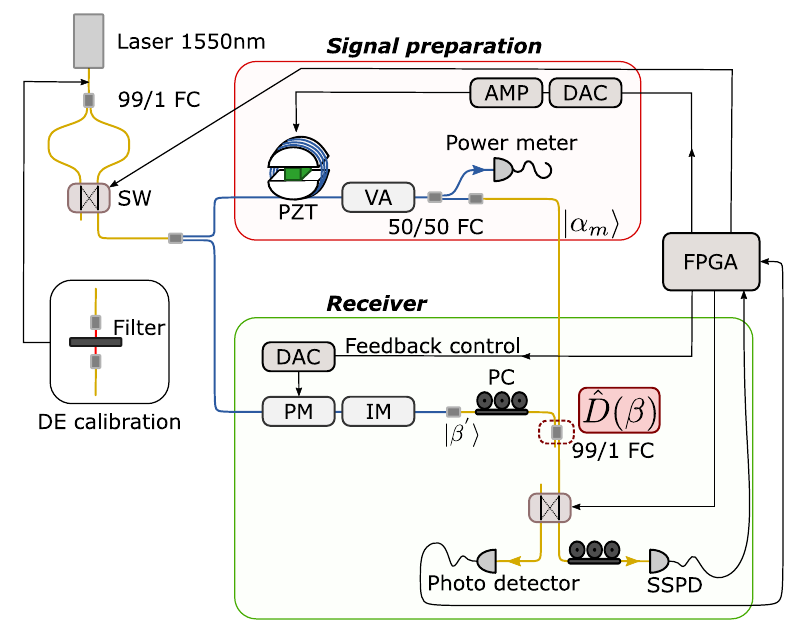}
}
\caption{
Experimental setup. Blue and yellow fibers respectively represent polarization maintaining fiber and single mode fiber.
FC: fiber coupler,
SW: switch,
PM: phase modulator,
IM: intensity modulator,
PZT: piezo transducer,
VA: variable attenuator,
PC: polarization controller,
SSPD: superconducting nanowire single photon detector,
DAC: digital to analog converter,
AMP: amplifier,
DE: detection efficiency.
}
\label{setup}
\end{figure}

Figure~\ref{setup} illustrates our experimental setup.
A continuous-wave, fiber-coupled laser at \SI{1550}{\nano\metre} is split in two paths, one for preparation of the signal states and one for the reference field for the displacement operation.
We prepare the QPSK signal states using a variable attenuator and a phase shifter that consists of a piezo transducer embedded in a circular mount with an optical fiber looped around, to control the intensity and the phase of the signal state respectively.
The displacement operation is physically implemented using a 99:1 fiber coupler, where the signal state interferes with the reference field for the displacement operation.
The amplitude of the displacement is controlled by an intensity modulator (operating with a constant voltage) that is calibrated to displace the signal states to the vacuum state by minimizing the count rate on a superconducting nanowire single photon detector (SSPD) \cite{SSPD,SSPD2}.
An interference visibility of $99.6 \%$ can be achieved by manually adjusting the polarization.
We switch the laser intensity between high and low using an optical switch in order to repeatedly and sequentially implement phase stabilization and measurement.
For the phase stabilization period, the relative phase between the signal and the reference is set to one of the four phase conditions $(2m+1)\pi/4$ $(m=0,1,2,3)$ by measuring the laser intensity after the 99:1 fiber coupler on a conventional photo detector (PD) and feeding back to the phase shifter. 
For the measurement, the displaced signal is detected by the SSPD.
In our experiment, instead of randomly preparing the QPSK coherent states, we prepare $500$ identical signal states after releasing the phase stabilization and repeat the procedure 20 times for each state.
A field programmable gate array (FPGA) counts the electrical signal from the SSPD and rapidly changes the voltages applied to the phase modulator for the reference field dependent on the counting history, which enables the feedback operation to the phase of the displacement. 
The feedback bandwidth of our receiver is about \SI{1}{\mega\hertz}, mainly limited by the speed of the digital to analog converter employed for the experiment.
Since the full time width of the signal state is defined to be $T$~=~\SI{200}{\micro\second}, the delay of the feedback is not negligible for our experiment and, hence, we discard the counts observed in the time interval $\Delta t$~=~\SI{1.1}{\micro\second} between each time bin.

Imperfect transmittance of optical components as well as the non-unit detection efficiency of the SSPD are the main sources of loss in our experiment.
The total transmittance from the 99:1 fiber coupler to the fiber right before the SSPD is measured to be $\eta_T\sim90\%$ (which is achieved by splicing the fibers) and the detection efficiency of the SSPD can reach $\eta_{\mathrm{SSPD}}\sim 73\%$ with a dark count rate of \SI{45}{\hertz}.
Therefore our receiver is expected to have a total system efficiency $\eta_{\mathrm{SE}}=\eta_T \times \eta_{\mathrm{SSPD}}$ of about $65\%$.
For the total system efficiency characterization, we insert free-space filters attenuating the laser power down to a few hundred thousand photons, which can be measured by the SSPD without saturation.
The laser intensity in the signal state preparation path is split into two paths using a 50/50 fiber coupler, where the laser power in one path is measured by a power meter to estimate the laser power $I_{\mathrm{in}}$ in the other path guided to the receiver.
After inserting a cascade of the calibrated optical filters at the beginning of the setup, the attenuated laser
passing through optical components of the receiver is detected by the SSPD.
Since we can estimate the photon rate from the laser power $I_{\mathrm{in}}$ and the attenuation of the inserted optical filters, 
the total system efficiency is obtained by comparing the photon rate and the observed count rate.
The total system efficiency can be evaluated with, at most, a $1.5\%$ uncertainty including the finite precision of calibrating the filters and the splitting ratio of the 50/50 fiber coupler, the systematic error due to the laser power fluctuation and the filter position.
As the detection efficiency of the SSPD depends on the polarization of the detected light, the polarization was carefully controlled to optimize the efficiency.

\begin{figure}[t]
\centering
{
\includegraphics[width=1\linewidth]
{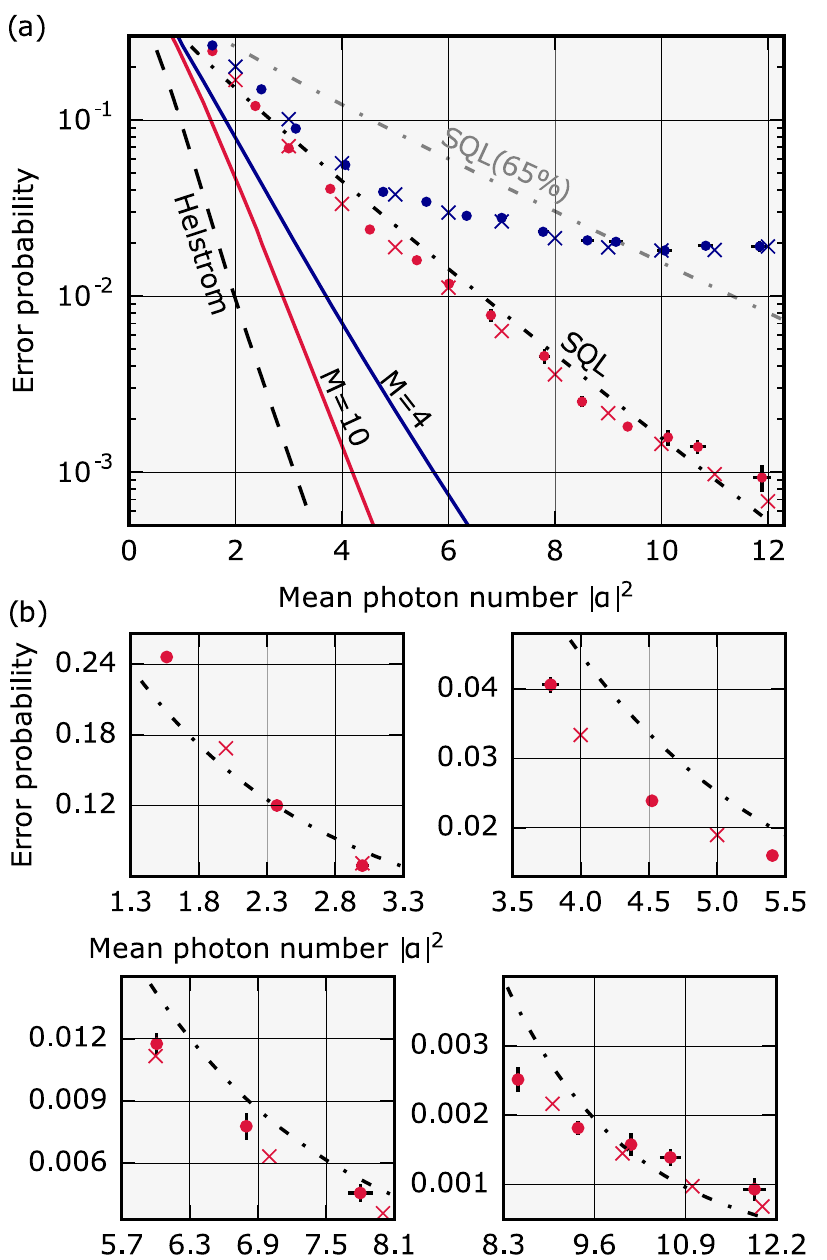}
}
\caption{
(a)
Error probability for the QPSK discrimination as a function of the signal mean photon number.
Red and blue points are experimentally obtained values for $M=10$ and $M=4$, and the corresponding theoretical predictions are shown by crosses. 
Red and blue solid lines are the performance of the feedback receiver with $M=10$ and $M=4$ in the ideal condition.
Black dashed, dashed-dot and thin dashed-dot lines represent the Helstrom bound, the SQL and the SQL with total detection efficiency 65$\%$.
(b)
The same data as in (a) but on a linear scale.
\label{Error}
}
\end{figure}

Experimentally obtained performance of our receiver is depicted in Figs.~\ref{Error}(a) and (b).
Blue and red dots represent the experimentally obtained error probability for $M=4$ and $M=10$ respectively.
The means and the error bars of the error probability are calculated from 5 independent procedures.
The signal mean photon number $\abs{\alpha}^2$ is estimated by rescaling the attenuated signal mean photon number $\eta_{\mathrm{SE}}\abs{\alpha}^2$ directly observed by the FPGA.
The error bar of the signal mean photon number originates from the finite precision of the estimation of the total system efficiency $\eta_{\mathrm{SE}}$.
Inconsistent size of the error bars on the signal mean photon number is due to the variation of the uncertainty of the total system efficiency associated with its characterization process. 
We measure the attenuated signal mean photon number before and after the measurement to ensure that the signal power is stable during the data acquisition and the difference of the attenuated signal mean photon numbers is typically less than $1\%$ of the attenuated signal power and $3\%$ at most.
As a conservative estimate of the mean photon number, we choose the largest of the two measurements for the plots in Figs.~\ref{Error}(a) and (b). 
The theoretical values shown by crosses are evaluated by Monte Carlo simulations under the experimental condition including the total system efficiency $\eta_{\mathrm{SE}}=65\%$, the visibility $\xi=99.6\%$,  the dark count noise $\nu=9.1\times 10^{-3}$ counts/state and a loss due to feedback delay compensation of $4.95\%$ for $M=10$ and $1.65\%$ for $M=4$.
Black dashed, dashed-dot and thin dashed-dot lines represent the Helstrom bound, the SQL and the SQL with total system efficiency 65$\%$.
The small deviation of the experimental results from the theoretical values is mainly due to variation of the visibility condition of the displacement.
For $M=4$,
the saturation of the error probability originates from the imperfection of the visibility and the dark count noise, which increase the probability of erroneously detecting the photon irrespective of the signal phase condition \cite{izumi2013,Becerra15}.
The effect of these phase insensitive counts is suppressed by performing a large number of feedback steps with the feedback strategy based on the {\it a posteriori} probability, appropriately taking into account these imperfections.
Although the performance of the receiver is far from ideal due to the finite detection efficiency and the visibility of the displacement operations, 
our receiver with $M=10$ outperforms the SQL and shows great agreement with the theoretical values.

\begin{figure}[t]
\centering
{
\includegraphics[width=1\linewidth]
{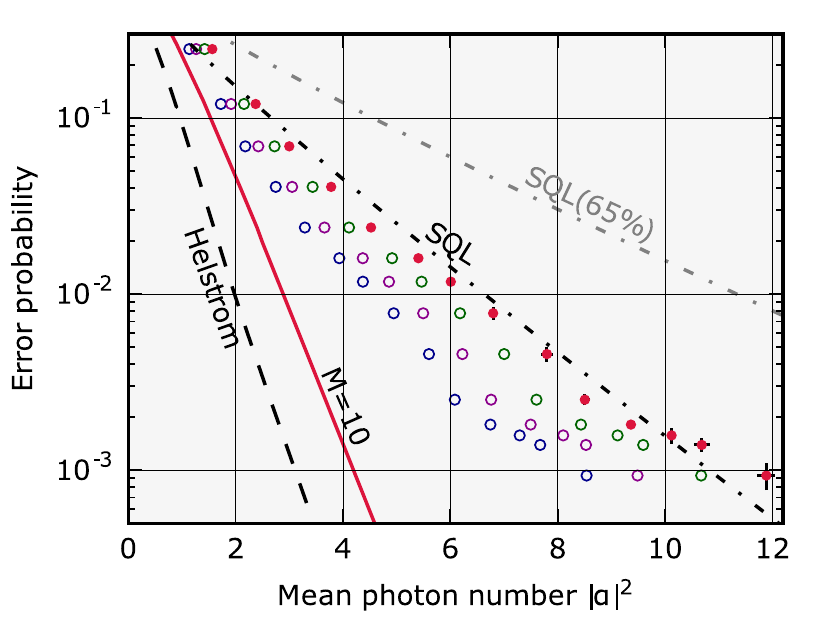}
}
\caption{
Expected error probabilities for various detection efficiencies of the SPD.
Blue, purple, green and red points are the expected error probabilities obtainable with the SPD with detection efficiencies 100$\%$, 90$\%$, 80$\%$ and 73$\%$, respectively.
\label{Error_varDE}
}
\end{figure}
We further investigate the possible performance of our receiver with an imaginary SPD assumed to have higher detection efficiencies.
In Fig.~\ref{Error_varDE}, we plot the expected performances of our receiver with an SPD whose efficiency is assumed to be $\eta_\mathrm{SSPD}$ = 100$\%$ (blue), 90$\%$ (purple), 80$\%$ (green) and finally 73$\%$ (red), corresponding to our experimental condition.
The signal mean photon number is obtained by rescaling the attenuated signal mean photon number $\eta_T\eta_{\mathrm{SSPD}}\abs{\alpha}^2$ with fixed $\eta_T = 90\%$.
Since the maximum detection efficiency of single photon detectors at telecom wavelength ever reported is above $90\%$ \cite{Lita:08,Marsili}, our receiver can potentially provide a significant improvement over the SQL by combining with such SPDs.
It is worth noting that the lowest requirement for the detection efficiency of the SPD to overcome the SQL using our system can be estimated to be $65\%$, corresponding to the total system efficiency $58.5\%$.
\begin{figure}[t]
\centering
{
\includegraphics[width=1 \linewidth]
{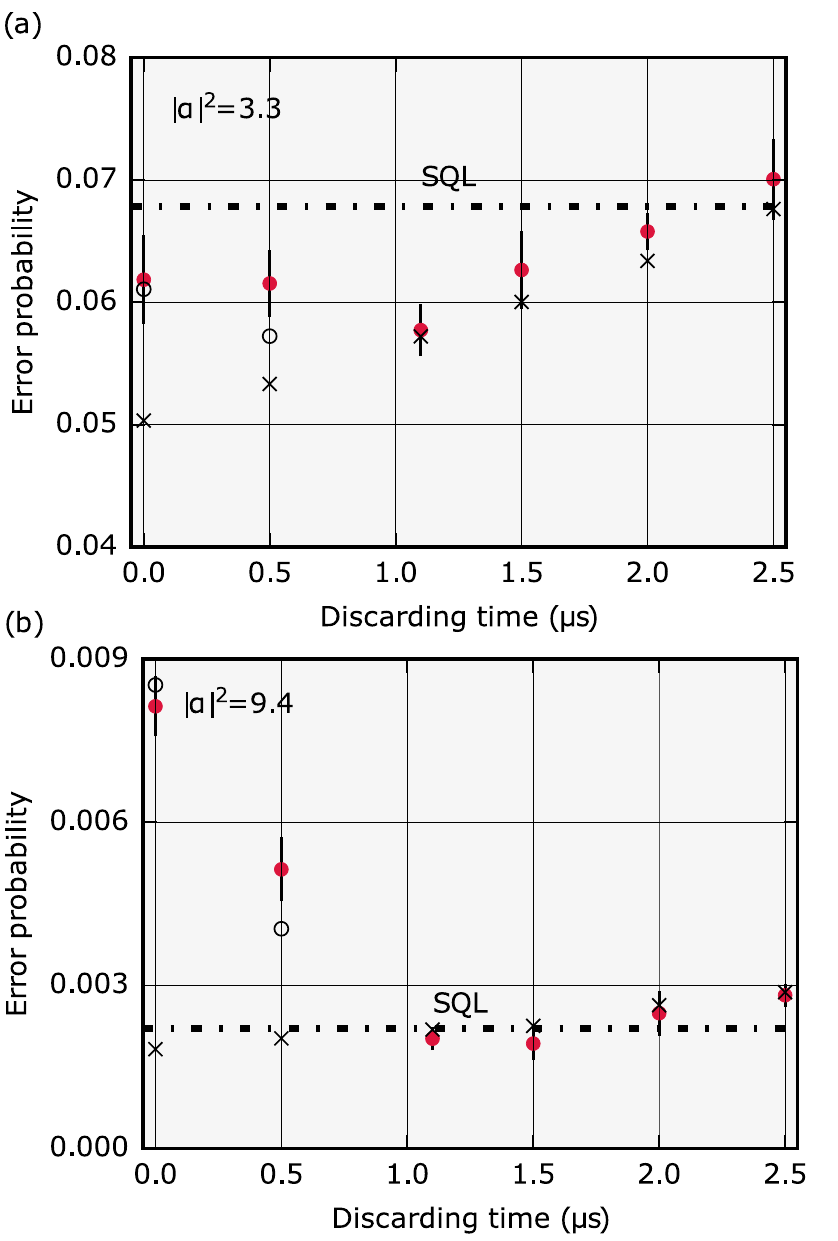}
}
\caption{
Error probability for QPSK discrimination with $M=10$ as a function of discarding time $\Delta t$.
(a) $\abs{\alpha}^2=3.3$, 
(b) $\abs{\alpha}^2=9.4$.
Red circles are the experimental results, while black circles and crosses are the theoretical values with and without the delay effect, respectively.
Special treatment of the delay effect is no longer necessary if the discarding time $\Delta t >$ \SI{1.0}{\micro\second} and the black circles are coincident with the black crosses.
Black dashed-dot line is the ideal SQL.
}
\label{Error_delay}
\end{figure}

Apart from the imperfections related to the photon detection, perhaps the most relevant practical imperfection of the feedback receiver is the limited bandwidth or delay of the feedback operation.
During the transition from one displacement phase to another, the displacement operation is not well-defined and spurious counts may occur, negatively affecting the performance of the receiver.
To alleviate the delay problem, the counts observed in a short time interval $\Delta t$ between each temporal mode bin can be discarded and the discarding time can be considered as additional linear loss.
We characterize this effect by setting the discarding time to various values and observing the error probability. We do this for mean photon numbers $\abs{\alpha}^2=3.3$ and $\abs{\alpha}^2=9.4$ and plot the results as a function of the discarding time $\Delta t$
in Figs.~\ref{Error_delay}(a) and (b).
The red circles represent the experimentally obtained error probabilities whose means and error bars are evaluated from 5 independent procedures.
The black circles and crosses are the theoretical values with and without the delay effect, obtained from the model discussed in Appendix.
For the black circles,
the delay is not considered in the probability distribution model to calculate the {\it a posteriori} probability for given outcomes, which is how our receiver was experimentally realized.
We could further improve the error probability if the delay effect is taken into account in the probability distribution model to calculate the {\it a posteriori} probability but we found that the improvement is not drastic and setting the discarding time to \SI{1.1}{\micro\second} shows a better performance anyway.
Since the displacement phase reaches the target condition within \SI{1.0}{\micro\second},
the theoretical values with the discarding time $\Delta t$ over \SI{1.0}{\micro\second} can be simply analyzed from the model with the discarding loss without the delay effect. The black circles are therefore coinciding with the black crosses.
For the simulation, we take into account the previously mentioned imperfections
as well as the discarding loss.
The degradation of the performance with small $\Delta t$ is due to the delay of the feedback operations and
the error probability can be improved by discarding the counts observed in $\Delta t$ but becomes higher as $\Delta t$ increases since the discarding loss increases.
Each data point is evaluated from independent experimental procedures and therefore the experimental conditions such as the mean photon number and the visibility of the displacement are slightly different in each discarding time condition. This explains the disagreement between the experiment and the theoretical values, which are calculated under the same parameter condition.
For small mean photon number, the delay effect is not as critical as for large photon number because the probability of detecting photons is low and the phase of the displacement operation is rarely changed.

\section{Conclusions}\label{Sect:4}
We experimentally realized a quantum receiver consisting of a displacement operation, a single photon detector and feedback operation. 
Our receiver employed a high performance photon counter and achieved discrimination of the QPSK signals with an error probability beating the SQL at telecom wavelength.
While the total system efficiency of our receiver is limited to $65\%$ mainly because of finite detection efficiency of the photon counter, our system could provide a substantial gain over the SQL by installing state-of-the-art photon counters showing higher performance at telecom wavelength \cite{Lita:08,Marsili}.
We further investigated the performance of the feedback measurement with finite bandwidth of the feedback.
Our delay analysis showed that the feedback delay could drastically degrade the performance for large signal mean photon numbers where the displacement operation needs to be more frequently changed. 

Since our all-fiber-based telecom quantum receiver is compatible with current optical fiber communication and can be modified to another type of two-dimensional projector such as a single-rail qubit projector \cite{PhysRevLett.124.070502}, we expect that our receiver will provide a practical advantage for not only conventional coherent communication but also quantum optical information processing \cite{Acin,ChavesBrask,Branciard}.
An interesting future step is to install our telecom quantum receiver in long distance fiber communication systems by developing a truly local reference field for the displacement operation \cite{Bina,PhysRevA.94.033842}.
Such a realistic communication scenario requires high speed transmission of signal states and therefore the bandwidth of the feedback operation would become a more critical challenge to be overcome.

Finally, further improvement of the feedback bandwidth is, in principle, possible by replacing the DAC with faster one. On the other hand, the ultimate feedback bandwidth of our system would be limited by a dead time of the SSPD, which is about \SI{50}{\nano\second}. One of the promising directions to increase the bandwidth would be installing a multi-pixcel SSPD that can effectively provide a short dead time \cite{doi:10.1063/1.4921318}.

\begin{acknowledgments}
This project was supported by Grant-in-Aid for JSPS Research Fellow, by VILLUM FONDEN via the Young Investigator Programme (Grant no. 10119) and by the Danish  National Research Foundation through the Center for Macroscopic Quantum States (bigQ DNRF142). 

\end{acknowledgments}

\appendix
%
\section{Delay analysis}
%
\begin{figure}[b]
\centering
{
\includegraphics[width=1 \linewidth]
{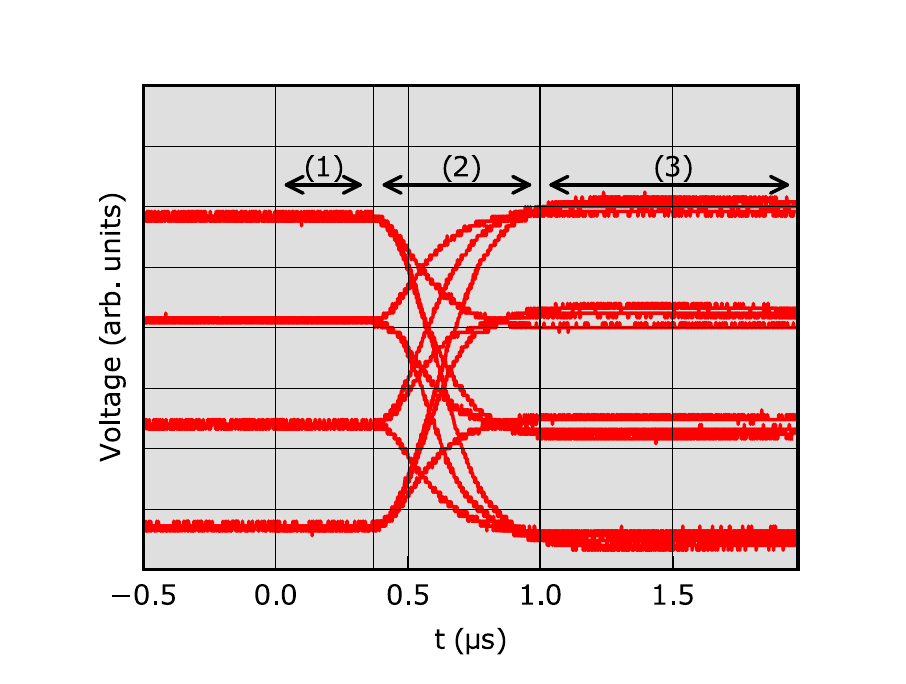}
}
\caption{
Electrical signals applied to the phase modulator.
The transition between time periods is at $t=0$.
The signal remains constant within the temporal region (1),
changes continuously to the target condition in (2),
and settles down to the target condition in (3).
}
\label{Signal_PM}
\end{figure}
Delay of feedback operations may degrade the performance of the receiver.
We develop a model of the receiver with finite bandwidth of the feedback operation to analyze the performance in the experimentally relevant condition.

We show the electrical signals applied to the phase modulator in Fig.~\ref{Signal_PM}.
The $j$'th time bin starts at $t=0$ and the electrical signal remains constant in (1), continuously changes to the target condition in (2), and settles down to the target condition in (3).
The delay of the electrical signal is because of the communication speed between the FPGA and the DAC in addition to the the finite response time of the DAC.
We employ a Nexys 4 artix-7 FPGA evaluation board and a Pmod DA3 with 16-bit resolution from Digilent.
\begin{figure}[t]
\centering
{
\includegraphics[width=1 \linewidth]
{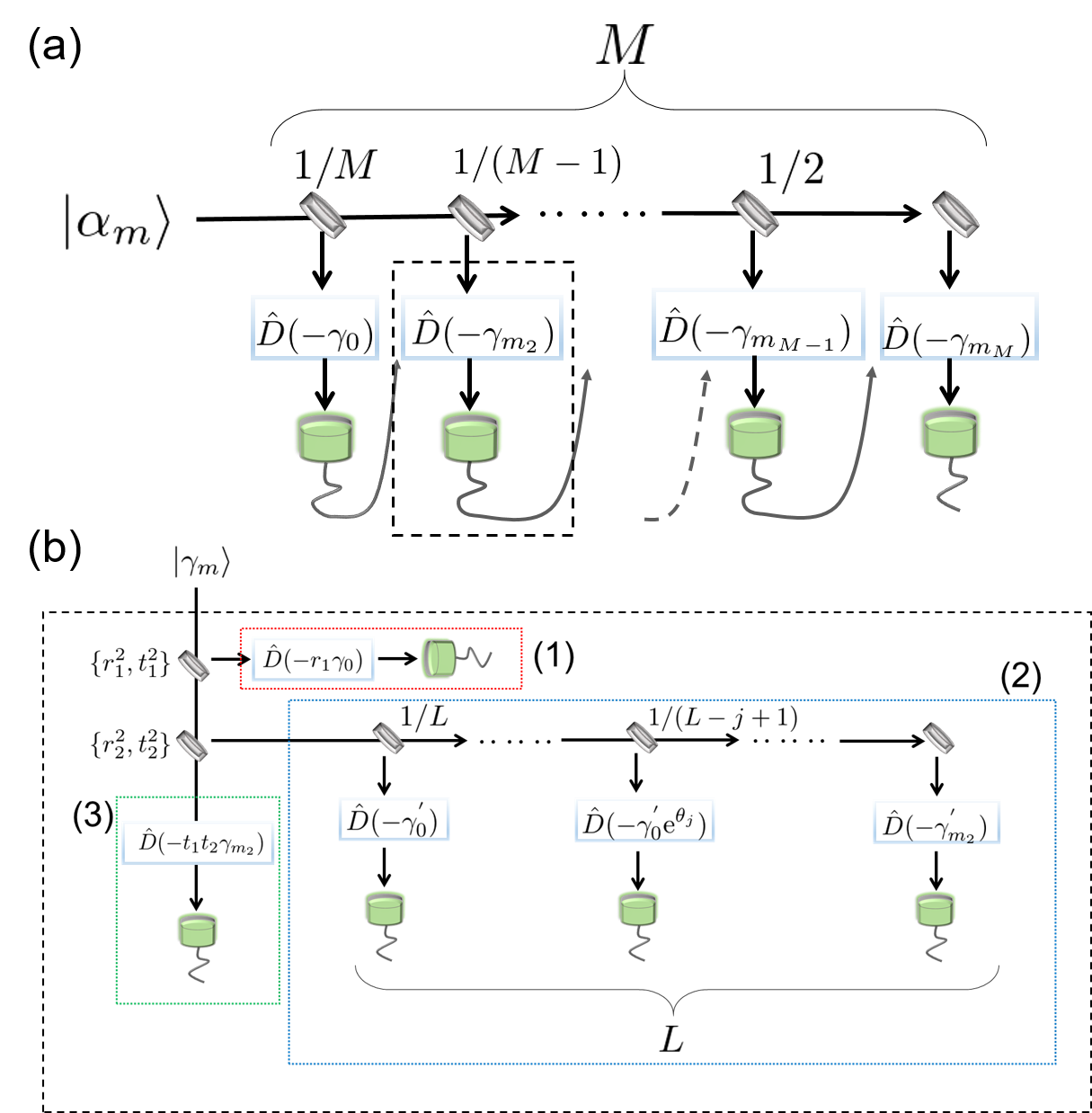}
}
\caption{
(a) Schematic of the feedback measurement with the SPDs in the spatial mode version.
(b) Theoretical model equivalent to the displacement condition with the delay shown in Fig.\ref{Signal_PM}. 
}
\label{Delay_model}
\end{figure}
In order to theoretically investigate the performance of the receiver with the delay of the feedback operation,
we adopt the spatial mode analysis where the signal state is spatially divided into $M$ modes and a displacement operation and a photon detection are performed on each mode.
Suppose that the signal state is equally split and the displacement operation is implemented such that one of the four signal states is displaced to the vacuum state $\hat{D}(-\gamma_m)\ket{\gamma_m}=\ket{0}$, where $\gamma_m=\alpha_m/\sqrt{M}$.
We focus on the second mode to discuss our model for the delay analysis.
In the second mode, the signal state is further divided by beam splitters with the reflectance (transmittance) $r_1^2$ ($t_1^2$) and $r_2^2$ ($t_2^2$) into three steps,
(1) the displacement operation remains to be the same as the previous displacement condition , (2) the displacement operation changes from the previous condition to the target condition, (3) the displacement operation is implemented in the target condition. 
A schematic of the model is shown in Fig.~\ref{Delay_model}(b).
The probability of having the ``off'' event in the second mode is given by a product of the ``off'' probabilities for all of the three steps.
For (1) and (3), the ``off'' probability can be simply given by,
\begin{eqnarray}
p(0|r_1(\gamma_{m}-\gamma_{0}))&=&
e^{-r_1^2\abs{\gamma_m-\gamma_{0}}^2}
\nonumber
\\
&=&
e^{-2r_1^2\gamma^2 (1-\xi\cos{\frac{m\pi}{2}})},
\nonumber
\\
p(0|t_1 t_2(\gamma_{m}-\gamma_{m_{2}}))&=&
e^{-t_1^2 t_2^2\abs{\gamma_m-\gamma_{m_{2}}}^2}
\nonumber
\\
&=&
e^{-2t_1^2 t_2^2\gamma^2 (1-\xi\cos{\frac{(m-m_2)\pi}{2}})},
\nonumber
\\
\label{Apeq1}
\end{eqnarray}
where $\xi$ is the visibility of the displacement.
In the step (2), where the displacement operation continuously changes from $\hat{D}(-\gamma_0)$ to $\hat{D}(-\gamma_{m_2})$,
we introduce the following model.
The signal state is equally split into $L$ modes and the displacement operation is first implemented such that $m=0$ is displaced to the vacuum state.
From the first mode to the $L$'th mode, the phase of the displacement operation is changed with the finite step $m_2\pi/2 \times 1/L$ , i.e., the displacement operation at $j$'th mode ($1\leq j \leq L$) is given by 
$\hat{D}(-\gamma_0^{'}e^{i \theta_j} )$, where $\gamma_m^{'}=t_1 r_2\gamma_m/\sqrt{L}$ and $\theta_j=m_2\pi/2 \times (j-1)/(L-1)$.
Therefore the probability of having the ``off'' event at $j$'th mode is, 
\begin{eqnarray}
p(0|\gamma_{m}^{'}-\gamma_{0}^{'}e^{i \theta_j})
&=&
e^{-\abs{\gamma_{m}^{'}-\gamma_{0}^{'}e^{i \theta_j}}^2}
\nonumber
\\
&=&
e^{-2\gamma^{'2} (1-\xi\cos{(\theta_j-\frac{m\pi}{2})})}.
\nonumber
\\
\label{Apeq2}
\end{eqnarray}
The `off'' event probability in second step (2) is given by a product of the ``off'' probabilities for $L$ modes,
\begin{eqnarray}
\prod_{j=1}^{L} p(0|\gamma_{m}^{'}-\gamma_{0}^{'}e^{i \theta_j})
&=&
e^{-\sum_{j=1}^{L}2\gamma^{'2} (1-\xi\cos{(\frac{m\pi}{2}-\theta_j)})},
\nonumber
\\
\label{Apeq3}
\end{eqnarray}
where
\begin{eqnarray}
&&\sum_{j=1}^{L}2\gamma^{'2} (1-\xi\cos{(\frac{m\pi}{2}-\theta_j)})
=
\nonumber
\\
&&
2t_1^2 r_2^2\gamma^{2}-\sum_{j=1}^{L} \frac{2t_1^2 r_2^2\gamma^{2}}{L} \xi\cos{(\frac{m\pi}{2}-\theta_j)}.
\label{Apeq4}
\end{eqnarray}
In order to analyze the continuously changing displacement,
we consider the limit of infinitely many modes $L\rightarrow\infty$,
\begin{eqnarray}
&&
\lim_{L\rightarrow \infty} \sum_{j=1}^{L} \frac{2t_1^2 r_2^2\gamma^{2}}{L} \xi\cos{(\frac{m\pi}{2}-\theta_j)}
=
\nonumber
\\
&&
\frac{4t_1^2 r_2^2\gamma^{2}}{m_2 \pi}\xi \bigl( \sin{\frac{m\pi}{2}}-\sin{\frac{(m-m_2)\pi}{2}}\bigr).
\label{Apeq5}
\end{eqnarray}
The probability of having the `off'' event in the second step (2) can be analytically obtained as,
\begin{eqnarray}
&&
\lim_{L\rightarrow \infty} \prod_{j=1}^{L} p(0|\gamma_{m}^{'}-\gamma_{0}^{'}e^{i \theta_j})
=
\nonumber
\\
&&
e^{-2t_1^2 r_2^2\gamma^{2}+\frac{4t_1^2 r_2^2\gamma^{2}}{m_2 \pi}\xi \bigl( \sin{\frac{m\pi}{2}}-\sin{\frac{(m-m_2)\pi}{2}}\bigr)}.
\label{Apeq6}
\end{eqnarray}
Finally, the probability of having the `off'' event for the second SPD in Fig.~\ref{Delay_model}(a) is obtained from a product of  `off'' event probabilities for three detection steps in Fig.~\ref{Delay_model}(b). 

Since we define a full time width of the signal state and each time bin as $200\,\mu$s and $20\,\mu$s for $M=10$, from Fig.~\ref{Signal_PM}, we find the parameters $r_1, r_2$ corresponding to our condition to be $r_1^2=0.37/20$ and $r_2^2=(1.0-0.37)/(20t_1^2)$.
The theoretical values for the performance of the receiver with the feedback delay shown in Fig.~\ref{Error_delay} are obtained from Monte Carlo simulation based on the model developed above. 
To simulate the discarding time $\Delta t=\SI{0.5}{\micro\second}$,
the first step (1) is discarded ($100\%$ linear loss), meaning that the outcome from the SPD is always `off'', and $21\%$ linear loss ((0.5-0.37)/(1.0-0.37)$\times$ 100) in the second step (2) is assumed.

\begin{figure}[t]
\centering
{
\includegraphics[width=1 \linewidth]
{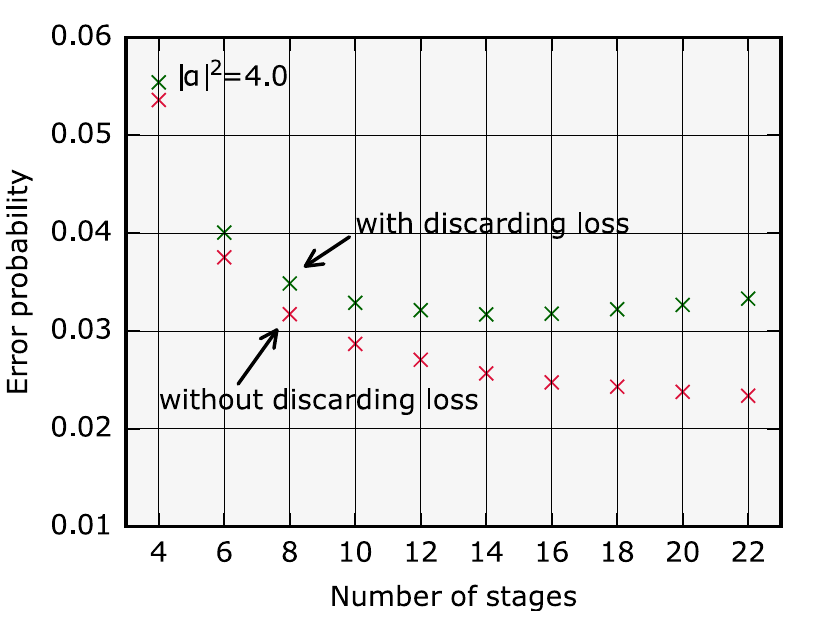}
}
\caption{
Error probability for the feedback measurement without the feedback discarding loss (red) and with the discarding loss (green) in the experimental condition, $\eta=65\%$, $\xi=99.6\%$ and $\nu=9.1\times 10^{-3}$ counts/state. The mean photon number of the signal state is set to $\abs{\alpha}^2=4.0$.
}
\label{Theory_error_delay}
\end{figure}

In our experiment, we set the discarding time $\Delta t=\SI{1.1}{\micro\second}$ and therefore the associated loss is $1.1/200\times (M-1)$.
Figure (\ref{Theory_error_delay}) shows the error probability of the feedback measurement for the mean photon number $\abs{\alpha}^2=4.0$ as a function of the number of feedback stages.
Red and green crosses respectively represent the simulation results for the feedback measurement under the experimental condition without the discarding loss assuming no delay of the feedback and with the discarding loss due to the delay compensation.
For an increasing number of stages, the performance eventually begins to degrade since the discarding loss becomes dominant.

%
\section{Simulation}
%
The theoretically expected values under the experimental condition are obtained by performing Monte Carlo simulation.
If the displacement magnitude is the same as the magnitude of the signal state,
the probability distribution of the photon detection can be described by a Poissonian distribution with the mean photon number $\bar{n}=\nu+2\eta (1-\xi\cos{\theta})\abs{\gamma}^2$, where $\theta$ is the relative phase between the signal state and the displacement direction \cite{doi:10.1080/09500340903203103,6600857,Becerra15}.
Hence, the probabilities of having the ``off'' event are given by
\begin{eqnarray}
&&
p(0|\gamma_m;-\gamma_m)=e^{-\nu-2\eta (1-\xi)\abs{\gamma}^2}
\nonumber
\\
&&
p(0|\gamma_{m+\frac{\pi}{2}};-\gamma_{m})=e^{-\nu-2\eta\abs{\gamma}^2}
\nonumber
\\
&&
p(0|\gamma_{m+\pi};-\gamma_{m})=e^{-\nu-2\eta (1+\xi)\abs{\gamma}^2}
\nonumber
\\
&&
p(0|\gamma_{m+\frac{3\pi}{2}};-\gamma_{m})=e^{-\nu-2\eta\abs{\gamma}^2},
\nonumber
\label{Apeq7}
\end{eqnarray}
where $p(0|\gamma_{m+\theta};-\gamma_{m})$ represents the probability of  the ``off'' event for the state whose phase is $\theta$ shifted from the target state displaced to the vacuum.
For the simulation, we first assume that the signal state $m=0$ is detected by the receiver and generate the outcome according to the Poissonian distribution with the mean photon number $\nu+2\eta (1-\xi)\abs{\gamma}^2$. 
Dependent on the outcome from the SPD, we calculate the {\it a posteriori} probabilities $P(\alpha_m|\{e_i\})$ for each state according to Eq.(\ref{eqposteriori}) and obtain the most-likely state maximizing the {\it a posteriori} probability. 
In the next step of the stages, the displacement operation is performed such that the most-likely state is displaced to the vacuum state, i.e., if the outcome is 0 corresponding to ``off'', the outcome is generated according to the Poissonian distribution with the same mean photon number and otherwise the Poissonian distribution with the mean photon number dependent on the {\it a posteriori} probabilities.
By repeating the simulation $M$ times and making the final decision based on the {\it a posteriori} probability at the last stage, we can conclude if we successfully identify the incoming signal or not.
We perform the similar procedure assuming that the signal states $m=1,2,3$ are respectively detected by the receiver with the same feedback strategy and the total error probability is calculated by summing up the error probabilities for each state divided by 4 ({\it a priori} probabilities).
We simulated the procedure $10^6$ times and obtained the average error probability for the QPSK discrimination with the feedback measurement.

The source code for the theoretical simulation of the strategies discussed in this paper is available at \cite{torefersuppref}.

\bibliography{reference}

\begin{thebibliography}{47}%
\makeatletter
\providecommand \@ifxundefined [1]{%
 \@ifx{#1\undefined}
}%
\providecommand \@ifnum [1]{%
 \ifnum #1\expandafter \@firstoftwo
 \else \expandafter \@secondoftwo
 \fi
}%
\providecommand \@ifx [1]{%
 \ifx #1\expandafter \@firstoftwo
 \else \expandafter \@secondoftwo
 \fi
}%
\providecommand \natexlab [1]{#1}%
\providecommand \enquote  [1]{``#1''}%
\providecommand \bibnamefont  [1]{#1}%
\providecommand \bibfnamefont [1]{#1}%
\providecommand \citenamefont [1]{#1}%
\providecommand \href@noop [0]{\@secondoftwo}%
\providecommand \href [0]{\begingroup \@sanitize@url \@href}%
\providecommand \@href[1]{\@@startlink{#1}\@@href}%
\providecommand \@@href[1]{\endgroup#1\@@endlink}%
\providecommand \@sanitize@url [0]{\catcode `\\12\catcode `\$12\catcode
  `\&12\catcode `\#12\catcode `\^12\catcode `\_12\catcode `\%12\relax}%
\providecommand \@@startlink[1]{}%
\providecommand \@@endlink[0]{}%
\providecommand \url  [0]{\begingroup\@sanitize@url \@url }%
\providecommand \@url [1]{\endgroup\@href {#1}{\urlprefix }}%
\providecommand \urlprefix  [0]{URL }%
\providecommand \Eprint [0]{\href }%
\providecommand \doibase [0]{https://doi.org/}%
\providecommand \selectlanguage [0]{\@gobble}%
\providecommand \bibinfo  [0]{\@secondoftwo}%
\providecommand \bibfield  [0]{\@secondoftwo}%
\providecommand \translation [1]{[#1]}%
\providecommand \BibitemOpen [0]{}%
\providecommand \bibitemStop [0]{}%
\providecommand \bibitemNoStop [0]{.\EOS\space}%
\providecommand \EOS [0]{\spacefactor3000\relax}%
\providecommand \BibitemShut  [1]{\csname bibitem#1\endcsname}%
\let\auto@bib@innerbib\@empty
\bibitem [{\citenamefont {Helstrom}(1976)}]{Helstrom_book76_QDET}%
  \BibitemOpen
  \bibfield  {author} {\bibinfo {author} {\bibfnamefont {C.~W.}\ \bibnamefont
  {Helstrom}},\ }\href@noop {} {\emph {\bibinfo {title} {Quantum Detection and
  Estimation Theory}}}\ (\bibinfo  {publisher} {Academic Press, New York},\
  \bibinfo {year} {1976})\BibitemShut {NoStop}%
\bibitem [{\citenamefont {Bennett}\ and\ \citenamefont
  {Brassard}(1984)}]{Bennett}%
  \BibitemOpen
  \bibfield  {author} {\bibinfo {author} {\bibfnamefont {C.~H.}\ \bibnamefont
  {Bennett}}\ and\ \bibinfo {author} {\bibfnamefont {G.}~\bibnamefont
  {Brassard}},\ }\bibfield  {title} {\bibinfo {title} {Quantum cryptography:
  Public key distribution and coin tossing},\ }in\ \href@noop {} {\emph
  {\bibinfo {booktitle} {Proc. of IEEE International Conference on Computers,
  Systems, and Signal Processing}}}\ (\bibinfo  {publisher} {IEEE, New York},\
  \bibinfo {year} {1984})\ pp.\ \bibinfo {pages} {175--179}\BibitemShut
  {NoStop}%
\bibitem [{\citenamefont {Gisin}\ \emph {et~al.}(2002)\citenamefont {Gisin},
  \citenamefont {Ribordy}, \citenamefont {Tittel},\ and\ \citenamefont
  {Zbinden}}]{RevModPhys.74.145}%
  \BibitemOpen
  \bibfield  {author} {\bibinfo {author} {\bibfnamefont {N.}~\bibnamefont
  {Gisin}}, \bibinfo {author} {\bibfnamefont {G.}~\bibnamefont {Ribordy}},
  \bibinfo {author} {\bibfnamefont {W.}~\bibnamefont {Tittel}},\ and\ \bibinfo
  {author} {\bibfnamefont {H.}~\bibnamefont {Zbinden}},\ }\bibfield  {title}
  {\bibinfo {title} {Quantum cryptography},\ }\href@noop {} {\bibfield
  {journal} {\bibinfo  {journal} {Rev. Mod. Phys.}\ }\textbf {\bibinfo {volume}
  {74}},\ \bibinfo {pages} {145} (\bibinfo {year} {2002})}\BibitemShut
  {NoStop}%
\bibitem [{\citenamefont {{Chan}}(2006)}]{4063386}%
  \BibitemOpen
  \bibfield  {author} {\bibinfo {author} {\bibfnamefont {V.~W.~S.}\
  \bibnamefont {{Chan}}},\ }\bibfield  {title} {\bibinfo {title} {Free-space
  optical communications},\ }\href@noop {} {\bibfield  {journal} {\bibinfo
  {journal} {Journal of Lightwave Technology}\ }\textbf {\bibinfo {volume}
  {24}},\ \bibinfo {pages} {4750} (\bibinfo {year} {2006})}\BibitemShut
  {NoStop}%
\bibitem [{\citenamefont {Liao}\ \emph {et~al.}(2017)\citenamefont {Liao},
  \citenamefont {Cai}, \citenamefont {Liu}, \citenamefont {Zhang},
  \citenamefont {Li}, \citenamefont {Ren}, \citenamefont {Yin}, \citenamefont
  {Shen}, \citenamefont {Cao}, \citenamefont {Li}, \citenamefont {Li},
  \citenamefont {Chen}, \citenamefont {Sun}, \citenamefont {Jia}, \citenamefont
  {Wu}, \citenamefont {Jiang}, \citenamefont {Wang}, \citenamefont {Huang},
  \citenamefont {Wang}, \citenamefont {Zhou}, \citenamefont {Deng},
  \citenamefont {Xi}, \citenamefont {Ma}, \citenamefont {Hu}, \citenamefont
  {Zhang}, \citenamefont {Chen}, \citenamefont {Liu}, \citenamefont {Wang},
  \citenamefont {Zhu}, \citenamefont {Lu}, \citenamefont {Shu}, \citenamefont
  {Peng}, \citenamefont {Wang},\ and\ \citenamefont {Pan}}]{LiaoQKD}%
  \BibitemOpen
  \bibfield  {author} {\bibinfo {author} {\bibfnamefont {S.-K.}\ \bibnamefont
  {Liao}}, \bibinfo {author} {\bibfnamefont {W.-Q.}\ \bibnamefont {Cai}},
  \bibinfo {author} {\bibfnamefont {W.-Y.}\ \bibnamefont {Liu}}, \bibinfo
  {author} {\bibfnamefont {L.}~\bibnamefont {Zhang}}, \bibinfo {author}
  {\bibfnamefont {Y.}~\bibnamefont {Li}}, \bibinfo {author} {\bibfnamefont
  {J.-G.}\ \bibnamefont {Ren}}, \bibinfo {author} {\bibfnamefont
  {J.}~\bibnamefont {Yin}}, \bibinfo {author} {\bibfnamefont {Q.}~\bibnamefont
  {Shen}}, \bibinfo {author} {\bibfnamefont {Y.}~\bibnamefont {Cao}}, \bibinfo
  {author} {\bibfnamefont {Z.-P.}\ \bibnamefont {Li}}, \bibinfo {author}
  {\bibfnamefont {F.-Z.}\ \bibnamefont {Li}}, \bibinfo {author} {\bibfnamefont
  {X.-W.}\ \bibnamefont {Chen}}, \bibinfo {author} {\bibfnamefont {L.-H.}\
  \bibnamefont {Sun}}, \bibinfo {author} {\bibfnamefont {J.-J.}\ \bibnamefont
  {Jia}}, \bibinfo {author} {\bibfnamefont {J.-C.}\ \bibnamefont {Wu}},
  \bibinfo {author} {\bibfnamefont {X.-J.}\ \bibnamefont {Jiang}}, \bibinfo
  {author} {\bibfnamefont {J.-F.}\ \bibnamefont {Wang}}, \bibinfo {author}
  {\bibfnamefont {Y.-M.}\ \bibnamefont {Huang}}, \bibinfo {author}
  {\bibfnamefont {Q.}~\bibnamefont {Wang}}, \bibinfo {author} {\bibfnamefont
  {Y.-L.}\ \bibnamefont {Zhou}}, \bibinfo {author} {\bibfnamefont
  {L.}~\bibnamefont {Deng}}, \bibinfo {author} {\bibfnamefont {T.}~\bibnamefont
  {Xi}}, \bibinfo {author} {\bibfnamefont {L.}~\bibnamefont {Ma}}, \bibinfo
  {author} {\bibfnamefont {T.}~\bibnamefont {Hu}}, \bibinfo {author}
  {\bibfnamefont {Q.}~\bibnamefont {Zhang}}, \bibinfo {author} {\bibfnamefont
  {Y.-A.}\ \bibnamefont {Chen}}, \bibinfo {author} {\bibfnamefont {N.-L.}\
  \bibnamefont {Liu}}, \bibinfo {author} {\bibfnamefont {X.-B.}\ \bibnamefont
  {Wang}}, \bibinfo {author} {\bibfnamefont {Z.-C.}\ \bibnamefont {Zhu}},
  \bibinfo {author} {\bibfnamefont {C.-Y.}\ \bibnamefont {Lu}}, \bibinfo
  {author} {\bibfnamefont {R.}~\bibnamefont {Shu}}, \bibinfo {author}
  {\bibfnamefont {C.-Z.}\ \bibnamefont {Peng}}, \bibinfo {author}
  {\bibfnamefont {J.-Y.}\ \bibnamefont {Wang}},\ and\ \bibinfo {author}
  {\bibfnamefont {J.-W.}\ \bibnamefont {Pan}},\ }\bibfield  {title} {\bibinfo
  {title} {Satellite-to-ground quantum key distribution},\ }\href@noop {}
  {\bibfield  {journal} {\bibinfo  {journal} {Nature}\ }\textbf {\bibinfo
  {volume} {549}},\ \bibinfo {pages} {43} (\bibinfo {year} {2017})}\BibitemShut
  {NoStop}%
\bibitem [{\citenamefont {Kikuchi}(2016)}]{7174950}%
  \BibitemOpen
  \bibfield  {author} {\bibinfo {author} {\bibfnamefont {K.}~\bibnamefont
  {Kikuchi}},\ }\bibfield  {title} {\bibinfo {title} {Fundamentals of coherent
  optical fiber communications},\ }\href@noop {} {\bibfield  {journal}
  {\bibinfo  {journal} {Journal of Lightwave Technology}\ }\textbf {\bibinfo
  {volume} {34}},\ \bibinfo {pages} {157} (\bibinfo {year} {2016})}\BibitemShut
  {NoStop}%
\bibitem [{\citenamefont {Takesue}\ \emph {et~al.}(2007)\citenamefont
  {Takesue}, \citenamefont {Nam}, \citenamefont {Zhang}, \citenamefont
  {Hadfield}, \citenamefont {Honjo}, \citenamefont {Tamaki},\ and\
  \citenamefont {Yamamoto}}]{TakesueQKD}%
  \BibitemOpen
  \bibfield  {author} {\bibinfo {author} {\bibfnamefont {H.}~\bibnamefont
  {Takesue}}, \bibinfo {author} {\bibfnamefont {S.~W.}\ \bibnamefont {Nam}},
  \bibinfo {author} {\bibfnamefont {Q.}~\bibnamefont {Zhang}}, \bibinfo
  {author} {\bibfnamefont {R.~H.}\ \bibnamefont {Hadfield}}, \bibinfo {author}
  {\bibfnamefont {T.}~\bibnamefont {Honjo}}, \bibinfo {author} {\bibfnamefont
  {K.}~\bibnamefont {Tamaki}},\ and\ \bibinfo {author} {\bibfnamefont
  {Y.}~\bibnamefont {Yamamoto}},\ }\bibfield  {title} {\bibinfo {title}
  {Quantum key distribution over a 40-d{B} channel loss using superconducting
  single-photon detectors},\ }\href@noop {} {\bibfield  {journal} {\bibinfo
  {journal} {Nature Photon.}\ }\textbf {\bibinfo {volume} {1}},\ \bibinfo
  {pages} {343} (\bibinfo {year} {2007})}\BibitemShut {NoStop}%
\bibitem [{\citenamefont {Caves}(1982)}]{PhysRevD.26.1817}%
  \BibitemOpen
  \bibfield  {author} {\bibinfo {author} {\bibfnamefont {C.~M.}\ \bibnamefont
  {Caves}},\ }\bibfield  {title} {\bibinfo {title} {Quantum limits on noise in
  linear amplifiers},\ }\href@noop {} {\bibfield  {journal} {\bibinfo
  {journal} {Phys. Rev. D}\ }\textbf {\bibinfo {volume} {26}},\ \bibinfo
  {pages} {1817} (\bibinfo {year} {1982})}\BibitemShut {NoStop}%
\bibitem [{\citenamefont {Tong}\ \emph {et~al.}(2011)\citenamefont {Tong},
  \citenamefont {Lundstr{\"o}m}, \citenamefont {Andrekson}, \citenamefont
  {McKinstrie}, \citenamefont {Karlsson}, \citenamefont {Blessing},
  \citenamefont {Tipsuwannakul}, \citenamefont {Puttnam}, \citenamefont
  {Toda},\ and\ \citenamefont {Gr{\"u}ner-Nielsen}}]{Tong2011}%
  \BibitemOpen
  \bibfield  {author} {\bibinfo {author} {\bibfnamefont {Z.}~\bibnamefont
  {Tong}}, \bibinfo {author} {\bibfnamefont {C.}~\bibnamefont {Lundstr{\"o}m}},
  \bibinfo {author} {\bibfnamefont {P.~A.}\ \bibnamefont {Andrekson}}, \bibinfo
  {author} {\bibfnamefont {C.~J.}\ \bibnamefont {McKinstrie}}, \bibinfo
  {author} {\bibfnamefont {M.}~\bibnamefont {Karlsson}}, \bibinfo {author}
  {\bibfnamefont {D.~J.}\ \bibnamefont {Blessing}}, \bibinfo {author}
  {\bibfnamefont {E.}~\bibnamefont {Tipsuwannakul}}, \bibinfo {author}
  {\bibfnamefont {B.~J.}\ \bibnamefont {Puttnam}}, \bibinfo {author}
  {\bibfnamefont {H.}~\bibnamefont {Toda}},\ and\ \bibinfo {author}
  {\bibfnamefont {L.}~\bibnamefont {Gr{\"u}ner-Nielsen}},\ }\bibfield  {title}
  {\bibinfo {title} {Towards ultrasensitive optical links enabled by low-noise
  phase-sensitive amplifiers},\ }\href@noop {} {\bibfield  {journal} {\bibinfo
  {journal} {Nature Photonics}\ }\textbf {\bibinfo {volume} {5}},\ \bibinfo
  {pages} {430} (\bibinfo {year} {2011})}\BibitemShut {NoStop}%
\bibitem [{\citenamefont {Asobe}\ \emph {et~al.}(2018)\citenamefont {Asobe},
  \citenamefont {Umeki},\ and\ \citenamefont {Tadanaga}}]{Asobe2018PhaseSA}%
  \BibitemOpen
  \bibfield  {author} {\bibinfo {author} {\bibfnamefont {M.}~\bibnamefont
  {Asobe}}, \bibinfo {author} {\bibfnamefont {T.}~\bibnamefont {Umeki}},\ and\
  \bibinfo {author} {\bibfnamefont {O.}~\bibnamefont {Tadanaga}},\ }\bibfield
  {title} {\bibinfo {title} {Phase sensitive amplifier using periodically poled
  {L}i{N}b{O}3 waveguides and their applications},\ }\href@noop {} {\bibfield
  {journal} {\bibinfo  {journal} {IEICE Transactions}\ }\textbf {\bibinfo
  {volume} {101-C}},\ \bibinfo {pages} {586} (\bibinfo {year}
  {2018})}\BibitemShut {NoStop}%
\bibitem [{\citenamefont {Ralph}\ \emph {et~al.}(2003)\citenamefont {Ralph},
  \citenamefont {Gilchrist}, \citenamefont {Milburn}, \citenamefont {Munro},\
  and\ \citenamefont {Glancy}}]{PhysRevA.68.042319}%
  \BibitemOpen
  \bibfield  {author} {\bibinfo {author} {\bibfnamefont {T.~C.}\ \bibnamefont
  {Ralph}}, \bibinfo {author} {\bibfnamefont {A.}~\bibnamefont {Gilchrist}},
  \bibinfo {author} {\bibfnamefont {G.~J.}\ \bibnamefont {Milburn}}, \bibinfo
  {author} {\bibfnamefont {W.~J.}\ \bibnamefont {Munro}},\ and\ \bibinfo
  {author} {\bibfnamefont {S.}~\bibnamefont {Glancy}},\ }\bibfield  {title}
  {\bibinfo {title} {Quantum computation with optical coherent states},\
  }\href@noop {} {\bibfield  {journal} {\bibinfo  {journal} {Phys. Rev. A}\
  }\textbf {\bibinfo {volume} {68}},\ \bibinfo {pages} {042319} (\bibinfo
  {year} {2003})}\BibitemShut {NoStop}%
\bibitem [{\citenamefont {{Eldar}}\ and\ \citenamefont
  {{Forney}}(2001)}]{915636}%
  \BibitemOpen
  \bibfield  {author} {\bibinfo {author} {\bibfnamefont {Y.~C.}\ \bibnamefont
  {{Eldar}}}\ and\ \bibinfo {author} {\bibfnamefont {G.~D.}\ \bibnamefont
  {{Forney}}},\ }\bibfield  {title} {\bibinfo {title} {On quantum detection and
  the square-root measurement},\ }\href@noop {} {\bibfield  {journal} {\bibinfo
   {journal} {IEEE Transactions on Information Theory}\ }\textbf {\bibinfo
  {volume} {47}},\ \bibinfo {pages} {858} (\bibinfo {year} {2001})}\BibitemShut
  {NoStop}%
\bibitem [{\citenamefont {Kennedy}(1973)}]{Kennedy73}%
  \BibitemOpen
  \bibfield  {author} {\bibinfo {author} {\bibfnamefont {R.~S.}\ \bibnamefont
  {Kennedy}},\ }\bibfield  {title} {\bibinfo {title} {A near-optimum receiver
  for the binary coherent state quantum channel},\ }\href@noop {} {\bibfield
  {journal} {\bibinfo  {journal} {Research Laboratory of Electronics, MIT,
  Quarterly Progress Report}\ ,\ \bibinfo {pages} {219}} (\bibinfo {year}
  {1973})}\BibitemShut {NoStop}%
\bibitem [{\citenamefont {Dolinar}(1973)}]{Dolinar73}%
  \BibitemOpen
  \bibfield  {author} {\bibinfo {author} {\bibfnamefont {S.}~\bibnamefont
  {Dolinar}},\ }\bibfield  {title} {\bibinfo {title} {An optimum receiver for
  the binary coherent state quantum channel},\ }\href@noop {} {\bibfield
  {journal} {\bibinfo  {journal} {Research Laboratory of Electronics, MIT,
  Quarterly Progress Report}\ ,\ \bibinfo {pages} {115}} (\bibinfo {year}
  {1973})}\BibitemShut {NoStop}%
\bibitem [{\citenamefont {Sasaki}\ and\ \citenamefont
  {Hirota}(1996)}]{PhysRevA.54.2728}%
  \BibitemOpen
  \bibfield  {author} {\bibinfo {author} {\bibfnamefont {M.}~\bibnamefont
  {Sasaki}}\ and\ \bibinfo {author} {\bibfnamefont {O.}~\bibnamefont
  {Hirota}},\ }\bibfield  {title} {\bibinfo {title} {Optimum decision scheme
  with a unitary control process for binary quantum-state signals},\
  }\href@noop {} {\bibfield  {journal} {\bibinfo  {journal} {Phys. Rev. A}\
  }\textbf {\bibinfo {volume} {54}},\ \bibinfo {pages} {2728} (\bibinfo {year}
  {1996})}\BibitemShut {NoStop}%
\bibitem [{\citenamefont {Takeoka}\ and\ \citenamefont
  {Sasaki}(2008)}]{PhysRevA.78.022320}%
  \BibitemOpen
  \bibfield  {author} {\bibinfo {author} {\bibfnamefont {M.}~\bibnamefont
  {Takeoka}}\ and\ \bibinfo {author} {\bibfnamefont {M.}~\bibnamefont
  {Sasaki}},\ }\bibfield  {title} {\bibinfo {title} {Discrimination of the
  binary coherent signal: Gaussian-operation limit and simple non-gaussian
  near-optimal receivers},\ }\href@noop {} {\bibfield  {journal} {\bibinfo
  {journal} {Phys. Rev. A}\ }\textbf {\bibinfo {volume} {78}},\ \bibinfo
  {pages} {022320} (\bibinfo {year} {2008})}\BibitemShut {NoStop}%
\bibitem [{\citenamefont {Wittmann}\ \emph {et~al.}(2008)\citenamefont
  {Wittmann}, \citenamefont {Takeoka}, \citenamefont {Cassemiro}, \citenamefont
  {Sasaki}, \citenamefont {Leuchs},\ and\ \citenamefont
  {Andersen}}]{PhysRevLett.101.210501}%
  \BibitemOpen
  \bibfield  {author} {\bibinfo {author} {\bibfnamefont {C.}~\bibnamefont
  {Wittmann}}, \bibinfo {author} {\bibfnamefont {M.}~\bibnamefont {Takeoka}},
  \bibinfo {author} {\bibfnamefont {K.~N.}\ \bibnamefont {Cassemiro}}, \bibinfo
  {author} {\bibfnamefont {M.}~\bibnamefont {Sasaki}}, \bibinfo {author}
  {\bibfnamefont {G.}~\bibnamefont {Leuchs}},\ and\ \bibinfo {author}
  {\bibfnamefont {U.~L.}\ \bibnamefont {Andersen}},\ }\bibfield  {title}
  {\bibinfo {title} {Demonstration of near-optimal discrimination of optical
  coherent states},\ }\href@noop {} {\bibfield  {journal} {\bibinfo  {journal}
  {Phys. Rev. Lett.}\ }\textbf {\bibinfo {volume} {101}},\ \bibinfo {pages}
  {210501} (\bibinfo {year} {2008})}\BibitemShut {NoStop}%
\bibitem [{\citenamefont {Tsujino}\ \emph {et~al.}(2010)\citenamefont
  {Tsujino}, \citenamefont {Fukuda}, \citenamefont {Fujii}, \citenamefont
  {Inoue}, \citenamefont {Fujiwara}, \citenamefont {Takeoka},\ and\
  \citenamefont {Sasaki}}]{Tsujino:10}%
  \BibitemOpen
  \bibfield  {author} {\bibinfo {author} {\bibfnamefont {K.}~\bibnamefont
  {Tsujino}}, \bibinfo {author} {\bibfnamefont {D.}~\bibnamefont {Fukuda}},
  \bibinfo {author} {\bibfnamefont {G.}~\bibnamefont {Fujii}}, \bibinfo
  {author} {\bibfnamefont {S.}~\bibnamefont {Inoue}}, \bibinfo {author}
  {\bibfnamefont {M.}~\bibnamefont {Fujiwara}}, \bibinfo {author}
  {\bibfnamefont {M.}~\bibnamefont {Takeoka}},\ and\ \bibinfo {author}
  {\bibfnamefont {M.}~\bibnamefont {Sasaki}},\ }\bibfield  {title} {\bibinfo
  {title} {Sub-shot-noise-limit discrimination of on-off keyed coherent signals
  via a quantum receiver with a superconducting transition edge sensor},\
  }\href@noop {} {\bibfield  {journal} {\bibinfo  {journal} {Opt. Express}\
  }\textbf {\bibinfo {volume} {18}},\ \bibinfo {pages} {8107} (\bibinfo {year}
  {2010})}\BibitemShut {NoStop}%
\bibitem [{\citenamefont {Tsujino}\ \emph {et~al.}(2011)\citenamefont
  {Tsujino}, \citenamefont {Fukuda}, \citenamefont {Fujii}, \citenamefont
  {Inoue}, \citenamefont {Fujiwara}, \citenamefont {Takeoka},\ and\
  \citenamefont {Sasaki}}]{PhysRevLett.106.250503}%
  \BibitemOpen
  \bibfield  {author} {\bibinfo {author} {\bibfnamefont {K.}~\bibnamefont
  {Tsujino}}, \bibinfo {author} {\bibfnamefont {D.}~\bibnamefont {Fukuda}},
  \bibinfo {author} {\bibfnamefont {G.}~\bibnamefont {Fujii}}, \bibinfo
  {author} {\bibfnamefont {S.}~\bibnamefont {Inoue}}, \bibinfo {author}
  {\bibfnamefont {M.}~\bibnamefont {Fujiwara}}, \bibinfo {author}
  {\bibfnamefont {M.}~\bibnamefont {Takeoka}},\ and\ \bibinfo {author}
  {\bibfnamefont {M.}~\bibnamefont {Sasaki}},\ }\bibfield  {title} {\bibinfo
  {title} {Quantum receiver beyond the standard quantum limit of coherent
  optical communication},\ }\href@noop {} {\bibfield  {journal} {\bibinfo
  {journal} {Phys. Rev. Lett.}\ }\textbf {\bibinfo {volume} {106}},\ \bibinfo
  {pages} {250503} (\bibinfo {year} {2011})}\BibitemShut {NoStop}%
\bibitem [{\citenamefont {DiMario}\ and\ \citenamefont
  {Becerra}(2018)}]{PhysRevLett.121.023603}%
  \BibitemOpen
  \bibfield  {author} {\bibinfo {author} {\bibfnamefont {M.~T.}\ \bibnamefont
  {DiMario}}\ and\ \bibinfo {author} {\bibfnamefont {F.~E.}\ \bibnamefont
  {Becerra}},\ }\bibfield  {title} {\bibinfo {title} {Robust measurement for
  the discrimination of binary coherent states},\ }\href@noop {} {\bibfield
  {journal} {\bibinfo  {journal} {Phys. Rev. Lett.}\ }\textbf {\bibinfo
  {volume} {121}},\ \bibinfo {pages} {023603} (\bibinfo {year}
  {2018})}\BibitemShut {NoStop}%
\bibitem [{\citenamefont {Cook}\ \emph {et~al.}(2007)\citenamefont {Cook},
  \citenamefont {Martin},\ and\ \citenamefont
  {Geremia}}]{CookMartinGeremia2007_Nature}%
  \BibitemOpen
  \bibfield  {author} {\bibinfo {author} {\bibfnamefont {R.~L.}\ \bibnamefont
  {Cook}}, \bibinfo {author} {\bibfnamefont {P.~J.}\ \bibnamefont {Martin}},\
  and\ \bibinfo {author} {\bibfnamefont {J.~M.}\ \bibnamefont {Geremia}},\
  }\bibfield  {title} {\bibinfo {title} {Optical coherent state discrimination
  using a closed-loop quantum measurement},\ }\href@noop {} {\bibfield
  {journal} {\bibinfo  {journal} {Nature}\ }\textbf {\bibinfo {volume} {446}},\
  \bibinfo {pages} {774} (\bibinfo {year} {2007})}\BibitemShut {NoStop}%
\bibitem [{\citenamefont {Takeoka}\ \emph {et~al.}(2005)\citenamefont
  {Takeoka}, \citenamefont {Sasaki}, \citenamefont {van Loock},\ and\
  \citenamefont {L\"utkenhaus}}]{TakeokaSasakiLutkenhaus2005}%
  \BibitemOpen
  \bibfield  {author} {\bibinfo {author} {\bibfnamefont {M.}~\bibnamefont
  {Takeoka}}, \bibinfo {author} {\bibfnamefont {M.}~\bibnamefont {Sasaki}},
  \bibinfo {author} {\bibfnamefont {P.}~\bibnamefont {van Loock}},\ and\
  \bibinfo {author} {\bibfnamefont {N.}~\bibnamefont {L\"utkenhaus}},\
  }\bibfield  {title} {\bibinfo {title} {Implementation of projective
  measurements with linear optics and continuous photon counting},\ }\href@noop
  {} {\bibfield  {journal} {\bibinfo  {journal} {Phys. Rev. A}\ }\textbf
  {\bibinfo {volume} {71}},\ \bibinfo {pages} {022318} (\bibinfo {year}
  {2005})}\BibitemShut {NoStop}%
\bibitem [{\citenamefont {Takeoka}\ \emph {et~al.}(2006)\citenamefont
  {Takeoka}, \citenamefont {Sasaki},\ and\ \citenamefont
  {L\"utkenhaus}}]{TakeokaSasakiLutkenhaus2006_PRL_BinaryProjMmt}%
  \BibitemOpen
  \bibfield  {author} {\bibinfo {author} {\bibfnamefont {M.}~\bibnamefont
  {Takeoka}}, \bibinfo {author} {\bibfnamefont {M.}~\bibnamefont {Sasaki}},\
  and\ \bibinfo {author} {\bibfnamefont {N.}~\bibnamefont {L\"utkenhaus}},\
  }\bibfield  {title} {\bibinfo {title} {Binary projective measurement via
  linear optics and photon counting},\ }\href@noop {} {\bibfield  {journal}
  {\bibinfo  {journal} {Phys. Rev. Lett.}\ }\textbf {\bibinfo {volume} {97}},\
  \bibinfo {pages} {040502} (\bibinfo {year} {2006})}\BibitemShut {NoStop}%
\bibitem [{\citenamefont {Izumi}\ \emph {et~al.}(2018)\citenamefont {Izumi},
  \citenamefont {Neergaard-Nielsen},\ and\ \citenamefont
  {Andersen}}]{Kennedytomography}%
  \BibitemOpen
  \bibfield  {author} {\bibinfo {author} {\bibfnamefont {S.}~\bibnamefont
  {Izumi}}, \bibinfo {author} {\bibfnamefont {J.~S.}\ \bibnamefont
  {Neergaard-Nielsen}},\ and\ \bibinfo {author} {\bibfnamefont {U.~L.}\
  \bibnamefont {Andersen}},\ }\bibfield  {title} {\bibinfo {title} {Tomography
  of a displacement photon counter for discrimination of single-rail optical
  qubits},\ }\href@noop {} {\bibfield  {journal} {\bibinfo  {journal} {Journal
  of Physics B: Atomic, Molecular and Optical Physics}\ }\textbf {\bibinfo
  {volume} {51}},\ \bibinfo {pages} {085502} (\bibinfo {year}
  {2018})}\BibitemShut {NoStop}%
\bibitem [{\citenamefont {Izumi}\ \emph {et~al.}(2020)\citenamefont {Izumi},
  \citenamefont {Neergaard-Nielsen},\ and\ \citenamefont
  {Andersen}}]{PhysRevLett.124.070502}%
  \BibitemOpen
  \bibfield  {author} {\bibinfo {author} {\bibfnamefont {S.}~\bibnamefont
  {Izumi}}, \bibinfo {author} {\bibfnamefont {J.~S.}\ \bibnamefont
  {Neergaard-Nielsen}},\ and\ \bibinfo {author} {\bibfnamefont {U.~L.}\
  \bibnamefont {Andersen}},\ }\bibfield  {title} {\bibinfo {title} {Tomography
  of a feedback measurement with photon detection},\ }\href@noop {} {\bibfield
  {journal} {\bibinfo  {journal} {Phys. Rev. Lett.}\ }\textbf {\bibinfo
  {volume} {124}},\ \bibinfo {pages} {070502} (\bibinfo {year}
  {2020})}\BibitemShut {NoStop}%
\bibitem [{\citenamefont {Bondurant}(1993)}]{Bondurant93}%
  \BibitemOpen
  \bibfield  {author} {\bibinfo {author} {\bibfnamefont {R.~S.}\ \bibnamefont
  {Bondurant}},\ }\bibfield  {title} {\bibinfo {title} {Near-quantum optimum
  receivers for the phase-quadrature coherent-state channel},\ }\href@noop {}
  {\bibfield  {journal} {\bibinfo  {journal} {Opt. Lett.}\ }\textbf {\bibinfo
  {volume} {18}},\ \bibinfo {pages} {1896} (\bibinfo {year}
  {1993})}\BibitemShut {NoStop}%
\bibitem [{\citenamefont {Müller}\ \emph {et~al.}(2012)\citenamefont
  {Müller}, \citenamefont {Usuga}, \citenamefont {Wittmann}, \citenamefont
  {Takeoka}, \citenamefont {Marquardt}, \citenamefont {Andersen},\ and\
  \citenamefont {Leuchs}}]{M_ller_2012}%
  \BibitemOpen
  \bibfield  {author} {\bibinfo {author} {\bibfnamefont {C.~R.}\ \bibnamefont
  {Müller}}, \bibinfo {author} {\bibfnamefont {M.~A.}\ \bibnamefont {Usuga}},
  \bibinfo {author} {\bibfnamefont {C.}~\bibnamefont {Wittmann}}, \bibinfo
  {author} {\bibfnamefont {M.}~\bibnamefont {Takeoka}}, \bibinfo {author}
  {\bibfnamefont {C.}~\bibnamefont {Marquardt}}, \bibinfo {author}
  {\bibfnamefont {U.~L.}\ \bibnamefont {Andersen}},\ and\ \bibinfo {author}
  {\bibfnamefont {G.}~\bibnamefont {Leuchs}},\ }\bibfield  {title} {\bibinfo
  {title} {Quadrature phase shift keying coherent state discrimination via a
  hybrid receiver},\ }\href@noop {} {\bibfield  {journal} {\bibinfo  {journal}
  {New Journal of Physics}\ }\textbf {\bibinfo {volume} {14}},\ \bibinfo
  {pages} {083009} (\bibinfo {year} {2012})}\BibitemShut {NoStop}%
\bibitem [{\citenamefont {Izumi}\ \emph {et~al.}(2012)\citenamefont {Izumi},
  \citenamefont {Takeoka}, \citenamefont {Fujiwara}, \citenamefont {Pozza},
  \citenamefont {Assalini}, \citenamefont {Ema},\ and\ \citenamefont
  {Sasaki}}]{izumi2012}%
  \BibitemOpen
  \bibfield  {author} {\bibinfo {author} {\bibfnamefont {S.}~\bibnamefont
  {Izumi}}, \bibinfo {author} {\bibfnamefont {M.}~\bibnamefont {Takeoka}},
  \bibinfo {author} {\bibfnamefont {M.}~\bibnamefont {Fujiwara}}, \bibinfo
  {author} {\bibfnamefont {N.~D.}\ \bibnamefont {Pozza}}, \bibinfo {author}
  {\bibfnamefont {A.}~\bibnamefont {Assalini}}, \bibinfo {author}
  {\bibfnamefont {K.}~\bibnamefont {Ema}},\ and\ \bibinfo {author}
  {\bibfnamefont {M.}~\bibnamefont {Sasaki}},\ }\bibfield  {title} {\bibinfo
  {title} {Displacement receiver for phase-shift-keyed coherent states},\
  }\href@noop {} {\bibfield  {journal} {\bibinfo  {journal} {Phys. Rev. A}\
  }\textbf {\bibinfo {volume} {86}},\ \bibinfo {pages} {042328} (\bibinfo
  {year} {2012})}\BibitemShut {NoStop}%
\bibitem [{\citenamefont {Izumi}\ \emph {et~al.}(2013)\citenamefont {Izumi},
  \citenamefont {Takeoka}, \citenamefont {Ema},\ and\ \citenamefont
  {Sasaki}}]{izumi2013}%
  \BibitemOpen
  \bibfield  {author} {\bibinfo {author} {\bibfnamefont {S.}~\bibnamefont
  {Izumi}}, \bibinfo {author} {\bibfnamefont {M.}~\bibnamefont {Takeoka}},
  \bibinfo {author} {\bibfnamefont {K.}~\bibnamefont {Ema}},\ and\ \bibinfo
  {author} {\bibfnamefont {M.}~\bibnamefont {Sasaki}},\ }\bibfield  {title}
  {\bibinfo {title} {Quantum receivers with squeezing and
  photon-number-resolving detectors for {$M$}-ary coherent state
  discrimination},\ }\href@noop {} {\bibfield  {journal} {\bibinfo  {journal}
  {Phys. Rev. A}\ }\textbf {\bibinfo {volume} {87}},\ \bibinfo {pages} {042328}
  (\bibinfo {year} {2013})}\BibitemShut {NoStop}%
\bibitem [{\citenamefont {Becerra}\ \emph {et~al.}(2013)\citenamefont
  {Becerra}, \citenamefont {Fan}, \citenamefont {Baumgartner}, \citenamefont
  {Goldhar}, \citenamefont {Kosloski},\ and\ \citenamefont
  {Migdall}}]{Becerra13}%
  \BibitemOpen
  \bibfield  {author} {\bibinfo {author} {\bibfnamefont {F.~E.}\ \bibnamefont
  {Becerra}}, \bibinfo {author} {\bibfnamefont {J.}~\bibnamefont {Fan}},
  \bibinfo {author} {\bibfnamefont {G.}~\bibnamefont {Baumgartner}}, \bibinfo
  {author} {\bibfnamefont {J.}~\bibnamefont {Goldhar}}, \bibinfo {author}
  {\bibfnamefont {J.~T.}\ \bibnamefont {Kosloski}},\ and\ \bibinfo {author}
  {\bibfnamefont {A.}~\bibnamefont {Migdall}},\ }\bibfield  {title} {\bibinfo
  {title} {Experimental demonstration of a receiver beating the standard
  quantum limit for multiple nonorthogonal state discrimination},\ }\href@noop
  {} {\bibfield  {journal} {\bibinfo  {journal} {Nature Photon.}\ }\textbf
  {\bibinfo {volume} {7}},\ \bibinfo {pages} {147} (\bibinfo {year}
  {2013})}\BibitemShut {NoStop}%
\bibitem [{\citenamefont {Becerra}\ \emph {et~al.}(2015)\citenamefont
  {Becerra}, \citenamefont {Fan},\ and\ \citenamefont {Migdall}}]{Becerra15}%
  \BibitemOpen
  \bibfield  {author} {\bibinfo {author} {\bibfnamefont {F.~E.}\ \bibnamefont
  {Becerra}}, \bibinfo {author} {\bibfnamefont {J.}~\bibnamefont {Fan}},\ and\
  \bibinfo {author} {\bibfnamefont {A.}~\bibnamefont {Migdall}},\ }\bibfield
  {title} {\bibinfo {title} {Photon number resolution enables quantum receiver
  for realistic coherent optical communications},\ }\href@noop {} {\bibfield
  {journal} {\bibinfo  {journal} {Nature Photon.}\ }\textbf {\bibinfo {volume}
  {9}},\ \bibinfo {pages} {48} (\bibinfo {year} {2015})}\BibitemShut {NoStop}%
\bibitem [{\citenamefont {Ferdinand}\ \emph {et~al.}(2017)\citenamefont
  {Ferdinand}, \citenamefont {DiMario},\ and\ \citenamefont
  {Becerra}}]{Ferdinand}%
  \BibitemOpen
  \bibfield  {author} {\bibinfo {author} {\bibfnamefont {A.~R.}\ \bibnamefont
  {Ferdinand}}, \bibinfo {author} {\bibfnamefont {M.~T.}\ \bibnamefont
  {DiMario}},\ and\ \bibinfo {author} {\bibfnamefont {F.~E.}\ \bibnamefont
  {Becerra}},\ }\bibfield  {title} {\bibinfo {title} {Multi-state
  discrimination below the quantum noise limit at the single-photon level},\
  }\href@noop {} {\bibfield  {journal} {\bibinfo  {journal} {npj Quantum
  Information}\ }\textbf {\bibinfo {volume} {3}},\ \bibinfo {pages} {43}
  (\bibinfo {year} {2017})}\BibitemShut {NoStop}%
\bibitem [{\citenamefont {{Li}}\ \emph {et~al.}(2013)\citenamefont {{Li}},
  \citenamefont {{Zuo}},\ and\ \citenamefont {{Zhu}}}]{6600857}%
  \BibitemOpen
  \bibfield  {author} {\bibinfo {author} {\bibfnamefont {K.}~\bibnamefont
  {{Li}}}, \bibinfo {author} {\bibfnamefont {Y.}~\bibnamefont {{Zuo}}},\ and\
  \bibinfo {author} {\bibfnamefont {B.}~\bibnamefont {{Zhu}}},\ }\bibfield
  {title} {\bibinfo {title} {Suppressing the errors due to mode mismatch for
  {$M$}-ary {PSK} quantum receivers using photon-number-resolving detector},\
  }\href@noop {} {\bibfield  {journal} {\bibinfo  {journal} {IEEE Photonics
  Technology Letters}\ }\textbf {\bibinfo {volume} {25}},\ \bibinfo {pages}
  {2182} (\bibinfo {year} {2013})}\BibitemShut {NoStop}%
\bibitem [{\citenamefont {Shcherbatenko}\ \emph {et~al.}(2020)\citenamefont
  {Shcherbatenko}, \citenamefont {Elezov}, \citenamefont {Goltsman},\ and\
  \citenamefont {Sych}}]{PhysRevA.101.032306}%
  \BibitemOpen
  \bibfield  {author} {\bibinfo {author} {\bibfnamefont {M.~L.}\ \bibnamefont
  {Shcherbatenko}}, \bibinfo {author} {\bibfnamefont {M.~S.}\ \bibnamefont
  {Elezov}}, \bibinfo {author} {\bibfnamefont {G.~N.}\ \bibnamefont
  {Goltsman}},\ and\ \bibinfo {author} {\bibfnamefont {D.~V.}\ \bibnamefont
  {Sych}},\ }\bibfield  {title} {\bibinfo {title} {Sub-shot-noise-limited
  fiber-optic quantum receiver},\ }\href@noop {} {\bibfield  {journal}
  {\bibinfo  {journal} {Phys. Rev. A}\ }\textbf {\bibinfo {volume} {101}},\
  \bibinfo {pages} {032306} (\bibinfo {year} {2020})}\BibitemShut {NoStop}%
\bibitem [{\citenamefont {Miki}\ \emph {et~al.}(2013)\citenamefont {Miki},
  \citenamefont {Yamashita}, \citenamefont {Terai},\ and\ \citenamefont
  {Wang}}]{SSPD}%
  \BibitemOpen
  \bibfield  {author} {\bibinfo {author} {\bibfnamefont {S.}~\bibnamefont
  {Miki}}, \bibinfo {author} {\bibfnamefont {T.}~\bibnamefont {Yamashita}},
  \bibinfo {author} {\bibfnamefont {H.}~\bibnamefont {Terai}},\ and\ \bibinfo
  {author} {\bibfnamefont {Z.}~\bibnamefont {Wang}},\ }\bibfield  {title}
  {\bibinfo {title} {High performance fiber-coupled {N}b{T}i{N} superconducting
  nanowire single photon detectors with {G}ifford-{M}c{M}ahon cryocooler},\
  }\href@noop {} {\bibfield  {journal} {\bibinfo  {journal} {Opt. Express}\
  }\textbf {\bibinfo {volume} {21}},\ \bibinfo {pages} {10208} (\bibinfo {year}
  {2013})}\BibitemShut {NoStop}%
\bibitem [{\citenamefont {Yamashita}\ \emph {et~al.}(2013)\citenamefont
  {Yamashita}, \citenamefont {Miki}, \citenamefont {Terai},\ and\ \citenamefont
  {Wang}}]{SSPD2}%
  \BibitemOpen
  \bibfield  {author} {\bibinfo {author} {\bibfnamefont {T.}~\bibnamefont
  {Yamashita}}, \bibinfo {author} {\bibfnamefont {S.}~\bibnamefont {Miki}},
  \bibinfo {author} {\bibfnamefont {H.}~\bibnamefont {Terai}},\ and\ \bibinfo
  {author} {\bibfnamefont {Z.}~\bibnamefont {Wang}},\ }\bibfield  {title}
  {\bibinfo {title} {Low-filling-factor superconducting single photon detector
  with high system detection efficiency},\ }\href@noop {} {\bibfield  {journal}
  {\bibinfo  {journal} {Opt. Express}\ }\textbf {\bibinfo {volume} {21}},\
  \bibinfo {pages} {27177} (\bibinfo {year} {2013})}\BibitemShut {NoStop}%
\bibitem [{\citenamefont {Marsili}\ \emph {et~al.}(2013)\citenamefont
  {Marsili}, \citenamefont {Verma}, \citenamefont {Stern}, \citenamefont
  {Harrington}, \citenamefont {Lita}, \citenamefont {Gerrits}, \citenamefont
  {Vayshenker}, \citenamefont {Baek}, \citenamefont {Shaw}, \citenamefont
  {Mirin},\ and\ \citenamefont {Nam}}]{Marsili}%
  \BibitemOpen
  \bibfield  {author} {\bibinfo {author} {\bibfnamefont {F.}~\bibnamefont
  {Marsili}}, \bibinfo {author} {\bibfnamefont {V.~B.}\ \bibnamefont {Verma}},
  \bibinfo {author} {\bibfnamefont {J.~A.}\ \bibnamefont {Stern}}, \bibinfo
  {author} {\bibfnamefont {S.}~\bibnamefont {Harrington}}, \bibinfo {author}
  {\bibfnamefont {A.~E.}\ \bibnamefont {Lita}}, \bibinfo {author}
  {\bibfnamefont {T.}~\bibnamefont {Gerrits}}, \bibinfo {author} {\bibfnamefont
  {I.}~\bibnamefont {Vayshenker}}, \bibinfo {author} {\bibfnamefont
  {B.}~\bibnamefont {Baek}}, \bibinfo {author} {\bibfnamefont {M.~D.}\
  \bibnamefont {Shaw}}, \bibinfo {author} {\bibfnamefont {R.~P.}\ \bibnamefont
  {Mirin}},\ and\ \bibinfo {author} {\bibfnamefont {S.~W.}\ \bibnamefont
  {Nam}},\ }\bibfield  {title} {\bibinfo {title} {Detecting single infrared
  photons with 93$\%$ system efficiency},\ }\href@noop {} {\bibfield  {journal}
  {\bibinfo  {journal} {Nature Photon.}\ }\textbf {\bibinfo {volume} {7}},\
  \bibinfo {pages} {210} (\bibinfo {year} {2013})}\BibitemShut {NoStop}%
\bibitem [{\citenamefont {Barnett}\ \emph {et~al.}(1998)\citenamefont
  {Barnett}, \citenamefont {Phillips},\ and\ \citenamefont
  {Pegg}}]{BARNETT199845}%
  \BibitemOpen
  \bibfield  {author} {\bibinfo {author} {\bibfnamefont {S.~M.}\ \bibnamefont
  {Barnett}}, \bibinfo {author} {\bibfnamefont {L.~S.}\ \bibnamefont
  {Phillips}},\ and\ \bibinfo {author} {\bibfnamefont {D.~T.}\ \bibnamefont
  {Pegg}},\ }\bibfield  {title} {\bibinfo {title} {Imperfect photodetection as
  projection onto mixed states},\ }\href@noop {} {\bibfield  {journal}
  {\bibinfo  {journal} {Optics Communications}\ }\textbf {\bibinfo {volume}
  {158}},\ \bibinfo {pages} {45 } (\bibinfo {year} {1998})}\BibitemShut
  {NoStop}%
\bibitem [{\citenamefont {Lita}\ \emph {et~al.}(2008)\citenamefont {Lita},
  \citenamefont {Miller},\ and\ \citenamefont {Nam}}]{Lita:08}%
  \BibitemOpen
  \bibfield  {author} {\bibinfo {author} {\bibfnamefont {A.~E.}\ \bibnamefont
  {Lita}}, \bibinfo {author} {\bibfnamefont {A.~J.}\ \bibnamefont {Miller}},\
  and\ \bibinfo {author} {\bibfnamefont {S.~W.}\ \bibnamefont {Nam}},\
  }\bibfield  {title} {\bibinfo {title} {Counting near-infrared single-photons
  with 95\% efficiency},\ }\href@noop {} {\bibfield  {journal} {\bibinfo
  {journal} {Opt. Express}\ }\textbf {\bibinfo {volume} {16}},\ \bibinfo
  {pages} {3032} (\bibinfo {year} {2008})}\BibitemShut {NoStop}%
\bibitem [{\citenamefont {Ac\'{\i}n}\ \emph {et~al.}(2007)\citenamefont
  {Ac\'{\i}n}, \citenamefont {Brunner}, \citenamefont {Gisin}, \citenamefont
  {Massar}, \citenamefont {Pironio},\ and\ \citenamefont {Scarani}}]{Acin}%
  \BibitemOpen
  \bibfield  {author} {\bibinfo {author} {\bibfnamefont {A.}~\bibnamefont
  {Ac\'{\i}n}}, \bibinfo {author} {\bibfnamefont {N.}~\bibnamefont {Brunner}},
  \bibinfo {author} {\bibfnamefont {N.}~\bibnamefont {Gisin}}, \bibinfo
  {author} {\bibfnamefont {S.}~\bibnamefont {Massar}}, \bibinfo {author}
  {\bibfnamefont {S.}~\bibnamefont {Pironio}},\ and\ \bibinfo {author}
  {\bibfnamefont {V.}~\bibnamefont {Scarani}},\ }\bibfield  {title} {\bibinfo
  {title} {Device-independent security of quantum cryptography against
  collective attacks},\ }\href@noop {} {\bibfield  {journal} {\bibinfo
  {journal} {Phys. Rev. Lett.}\ }\textbf {\bibinfo {volume} {98}},\ \bibinfo
  {pages} {230501} (\bibinfo {year} {2007})}\BibitemShut {NoStop}%
\bibitem [{\citenamefont {Chaves}\ and\ \citenamefont
  {Brask}(2011)}]{ChavesBrask}%
  \BibitemOpen
  \bibfield  {author} {\bibinfo {author} {\bibfnamefont {R.}~\bibnamefont
  {Chaves}}\ and\ \bibinfo {author} {\bibfnamefont {J.~B.}\ \bibnamefont
  {Brask}},\ }\bibfield  {title} {\bibinfo {title} {Feasibility of
  loophole-free nonlocality tests with a single photon},\ }\href@noop {}
  {\bibfield  {journal} {\bibinfo  {journal} {Phys. Rev. A}\ }\textbf {\bibinfo
  {volume} {84}},\ \bibinfo {pages} {062110} (\bibinfo {year}
  {2011})}\BibitemShut {NoStop}%
\bibitem [{\citenamefont {Branciard}\ \emph {et~al.}(2012)\citenamefont
  {Branciard}, \citenamefont {Cavalcanti}, \citenamefont {Walborn},
  \citenamefont {Scarani},\ and\ \citenamefont {Wiseman}}]{Branciard}%
  \BibitemOpen
  \bibfield  {author} {\bibinfo {author} {\bibfnamefont {C.}~\bibnamefont
  {Branciard}}, \bibinfo {author} {\bibfnamefont {E.~G.}\ \bibnamefont
  {Cavalcanti}}, \bibinfo {author} {\bibfnamefont {S.~P.}\ \bibnamefont
  {Walborn}}, \bibinfo {author} {\bibfnamefont {V.}~\bibnamefont {Scarani}},\
  and\ \bibinfo {author} {\bibfnamefont {H.~M.}\ \bibnamefont {Wiseman}},\
  }\bibfield  {title} {\bibinfo {title} {One-sided device-independent quantum
  key distribution: Security, feasibility, and the connection with steering},\
  }\href@noop {} {\bibfield  {journal} {\bibinfo  {journal} {Phys. Rev. A}\
  }\textbf {\bibinfo {volume} {85}},\ \bibinfo {pages} {010301} (\bibinfo
  {year} {2012})}\BibitemShut {NoStop}%
\bibitem [{\citenamefont {Bina}\ \emph {et~al.}(2016)\citenamefont {Bina},
  \citenamefont {Allevi}, \citenamefont {Bondani},\ and\ \citenamefont
  {Olivares}}]{Bina}%
  \BibitemOpen
  \bibfield  {author} {\bibinfo {author} {\bibfnamefont {M.}~\bibnamefont
  {Bina}}, \bibinfo {author} {\bibfnamefont {A.}~\bibnamefont {Allevi}},
  \bibinfo {author} {\bibfnamefont {M.}~\bibnamefont {Bondani}},\ and\ \bibinfo
  {author} {\bibfnamefont {S.}~\bibnamefont {Olivares}},\ }\bibfield  {title}
  {\bibinfo {title} {Phase-reference monitoring in coherent-state
  discrimination assisted by a photon-number resolving detector},\ }\href@noop
  {} {\bibfield  {journal} {\bibinfo  {journal} {Sci. Rep.}\ }\textbf {\bibinfo
  {volume} {6}},\ \bibinfo {pages} {26025} (\bibinfo {year}
  {2016})}\BibitemShut {NoStop}%
\bibitem [{\citenamefont {Izumi}\ \emph {et~al.}(2016)\citenamefont {Izumi},
  \citenamefont {Takeoka}, \citenamefont {Wakui}, \citenamefont {Fujiwara},
  \citenamefont {Ema},\ and\ \citenamefont {Sasaki}}]{PhysRevA.94.033842}%
  \BibitemOpen
  \bibfield  {author} {\bibinfo {author} {\bibfnamefont {S.}~\bibnamefont
  {Izumi}}, \bibinfo {author} {\bibfnamefont {M.}~\bibnamefont {Takeoka}},
  \bibinfo {author} {\bibfnamefont {K.}~\bibnamefont {Wakui}}, \bibinfo
  {author} {\bibfnamefont {M.}~\bibnamefont {Fujiwara}}, \bibinfo {author}
  {\bibfnamefont {K.}~\bibnamefont {Ema}},\ and\ \bibinfo {author}
  {\bibfnamefont {M.}~\bibnamefont {Sasaki}},\ }\bibfield  {title} {\bibinfo
  {title} {Optical phase estimation via the coherent state and displaced-photon
  counting},\ }\href@noop {} {\bibfield  {journal} {\bibinfo  {journal} {Phys.
  Rev. A}\ }\textbf {\bibinfo {volume} {94}},\ \bibinfo {pages} {033842}
  (\bibinfo {year} {2016})}\BibitemShut {NoStop}%
\bibitem [{\citenamefont {Allman}\ \emph {et~al.}(2015)\citenamefont {Allman},
  \citenamefont {Verma}, \citenamefont {Stevens}, \citenamefont {Gerrits},
  \citenamefont {Horansky}, \citenamefont {Lita}, \citenamefont {Marsili},
  \citenamefont {Beyer}, \citenamefont {Shaw}, \citenamefont {Kumor},
  \citenamefont {Mirin},\ and\ \citenamefont {Nam}}]{doi:10.1063/1.4921318}%
  \BibitemOpen
  \bibfield  {author} {\bibinfo {author} {\bibfnamefont {M.~S.}\ \bibnamefont
  {Allman}}, \bibinfo {author} {\bibfnamefont {V.~B.}\ \bibnamefont {Verma}},
  \bibinfo {author} {\bibfnamefont {M.}~\bibnamefont {Stevens}}, \bibinfo
  {author} {\bibfnamefont {T.}~\bibnamefont {Gerrits}}, \bibinfo {author}
  {\bibfnamefont {R.~D.}\ \bibnamefont {Horansky}}, \bibinfo {author}
  {\bibfnamefont {A.~E.}\ \bibnamefont {Lita}}, \bibinfo {author}
  {\bibfnamefont {F.}~\bibnamefont {Marsili}}, \bibinfo {author} {\bibfnamefont
  {A.}~\bibnamefont {Beyer}}, \bibinfo {author} {\bibfnamefont {M.~D.}\
  \bibnamefont {Shaw}}, \bibinfo {author} {\bibfnamefont {D.}~\bibnamefont
  {Kumor}}, \bibinfo {author} {\bibfnamefont {R.}~\bibnamefont {Mirin}},\ and\
  \bibinfo {author} {\bibfnamefont {S.~W.}\ \bibnamefont {Nam}},\ }\bibfield
  {title} {\bibinfo {title} {A near-infrared 64-pixel superconducting nanowire
  single photon detector array with integrated multiplexed readout},\
  }\href@noop {} {\bibfield  {journal} {\bibinfo  {journal} {Applied Physics
  Letters}\ }\textbf {\bibinfo {volume} {106}},\ \bibinfo {pages} {192601}
  (\bibinfo {year} {2015})}\BibitemShut {NoStop}%
\bibitem [{\citenamefont {Takeoka}\ \emph {et~al.}(2010)\citenamefont
  {Takeoka}, \citenamefont {Tsujino},\ and\ \citenamefont
  {Sasaki}}]{doi:10.1080/09500340903203103}%
  \BibitemOpen
  \bibfield  {author} {\bibinfo {author} {\bibfnamefont {M.}~\bibnamefont
  {Takeoka}}, \bibinfo {author} {\bibfnamefont {K.}~\bibnamefont {Tsujino}},\
  and\ \bibinfo {author} {\bibfnamefont {M.}~\bibnamefont {Sasaki}},\
  }\bibfield  {title} {\bibinfo {title} {Cut-off rate analysis of practical
  quantum receivers},\ }\href@noop {} {\bibfield  {journal} {\bibinfo
  {journal} {Journal of Modern Optics}\ }\textbf {\bibinfo {volume} {57}},\
  \bibinfo {pages} {207} (\bibinfo {year} {2010})}\BibitemShut {NoStop}%
\bibitem [{tor()}]{torefersuppref}%
  \BibitemOpen
  \href@noop {} {}\bibinfo {howpublished}
  {\url{https://github.com/qpit/QPSK_discrimination.git}}\BibitemShut {NoStop}%
\end{thebibliography}%

\end{document}